\documentclass[12pt]{iopart}

\usepackage{graphicx}
\usepackage{amsthm}
\usepackage{iopams}
\usepackage{color}
\usepackage{bbm}
\usepackage{ulem}
\usepackage[dvipsnames]{xcolor}

\newcommand{\ket}[1]{\mbox{$ | #1 \rangle $}}
\newcommand{\bra}[1]{\mbox{$ \langle #1 | $}}

\begin{document}

\title{Decoy-state quantum key distribution with a leaky source}

\author{Kiyoshi Tamaki$^{1,*,\dagger}$, Marcos Curty$^{2,*}$ and Marco Lucamarini$^{3,4,*}$}
\address{$^1$ NTT Basic Research Laboratories, NTT Corporation, 3-1, Morinosato-Wakamiya Atsugi-Shi, 243-0198, Japan}
\address{$^2$ EI Telecomunicaci\'on, Department of Signal Theory and Communications, University of Vigo, Vigo E-36310, Spain}
\address{$^3$ Toshiba Research Europe Ltd, 208 Cambridge Science Park, Cambridge CB4 0GZ, United Kingdom}
\address{$^4$ Toshiba Corporate Research \& Development Center, 1 Komukai-Toshiba-Cho, Saiwai-ku, Kawasaki 212-8582, Japan}
\address{$^*$ All authors contributed equally to this work\\
$\dagger$ Currently his affiliation is: Graduate School of Science and Engineering for Education,
University of Toyama, Gofuku 3190, Toyama 930-8555, Japan}

\eads{\mailto{mcurty@com.uvigo.es}}

\begin{abstract}
In recent years, there has been a great effort to prove the security of quantum key distribution (QKD) with a minimum number of assumptions. Besides its intrinsic theoretical interest, this would allow for larger tolerance against device imperfections in the actual implementations. However, even in this device-independent scenario, one assumption seems unavoidable, that is, the presence of a protected space devoid of any unwanted information leakage in which the legitimate parties can privately generate, process and store their classical data.
In this paper we relax this unrealistic and hardly feasible assumption and introduce a general formalism to tackle the information leakage problem in most of existing QKD systems. More specifically, we prove the security of optical QKD systems using phase and intensity modulators in their transmitters, which leak the setting information in an arbitrary manner. We apply our security proof to cases of practical interest and show key rates similar to those obtained in a perfectly shielded environment. Our work constitutes a fundamental step forward in guaranteeing implementation security of quantum communication systems.
\end{abstract}

\section{Introduction}

It is well-known that two spatially separated users (Alice and Bob) can secretly communicate over a public channel if they own two identical random keys unknown to any third party. They can use their keys to enable symmetric-key encryption. When the symmetric-key algorithm is the so-called ``one-time pad''~\cite{vernam}, the security of the resulting communication is independent of the computational capability of an eavesdropper (Eve)~\cite{shannon}. The only provably secure way known to date to distill secret random keys at remote locations is quantum key distribution~(QKD)~\cite{qkd1,qkd2,qkd3,qkd4}. While the theoretical security of QKD has been convincingly proven in recent years~\cite{qkd3}, in practice a QKD realisation cannot typically perfectly satisfy the requirements imposed by the theory. Therefore it is crucial that security proofs are extended to accommodate the imperfections of the real QKD devices. Any unaccounted imperfection constitutes a so-called ``side-channel'', which can be exploited by Eve to compromise the security of the system~\cite{sc1,sc2,sc3,sc4,sc5,sc6,sc7,sc8,sc9,sc10,sc11}.

To close the gap between theory and practice, various approaches have been proposed so far, with two most prominent examples being ``device-independent QKD''~\cite{diQKD1,diQKD2,diQKD3,diQKD4} and decoy-state ``measurement-device-independent QKD'' (mdiQKD)~\cite{mdiQKD}. Device-independent QKD does not require a complete knowledge of how QKD apparatuses operate, being its security based on the violation of a Bell inequality. However, its experimental complexity is unsuitable for practical applications, as its ultimate form demands that Alice and Bob perform a loophole-free Bell test~\cite{belltest1,belltest2,belltest3} in every QKD session. Also, its secret key rate is very poor with current technology~\cite{amp1,amp2}. Decoy-state mdiQKD, on the other hand, permits to remove any assumption of trustfulness from the measurement device, which is arguably the weakest part of QKD realisations~\cite{sc1,sc2,sc3,sc4,sc5,sc6,sc7,sc8}. Under the only additional requirement that Alice and Bob know their state preparation process~\cite{loss_t}, mdiQKD with decoy-states allows to bring QKD theory closer to practice~\cite{review_mdiQKD} without frustrating the key rate~\cite{mdiQKD,mdiQKDfinite}.
Most importantly, its practical feasibility has been already experimentally demonstrated both in laboratories and in field trials~\cite{expmdi1,expmdi2,expmdi3,expmdi4,expmdi5,expmdi6,expmdi7,expmdi8}, with a key rate comparable to that of standard QKD protocols~\cite{expmdi7}.

However, it is important to notice that the security of any form of QKD, including the two solutions above, relies on the assumption that Alice and Bob's devices do not leak any unwanted information to the outside. That is, their apparatuses must be inside private spaces that are well-shielded and inaccessible to Eve (see, e.g., \cite{BCK13}).
This assumption is very hard, if not impossible, to guarantee in practice.
The behaviour of real devices is affected by the environmental conditions and can depend on their response to external signals, unawarely triggered by a legitimate user, or maliciously injected into the QKD system by Eve. This could open new side-channels, of which the so-called Trojan-Horse attack (THA)~\cite{tha1,tha2,tha3} is a meaningful example.
While mdiQKD relieves QKD from the burden of characterising the measuring devices, the THA deals with the important question of guaranteeing a protected boundary between the transmitting devices, assigned with the preparation of the initial quantum states, and the outside world.

In a THA, Eve injects bright light pulses into the users' devices and analyse the back-reflected light, with the aim of extracting more information from the signals travelling in the quantum channel. Recently,~\cite{tha3} considered a feasible THA targeting the phase modulator (PM) of a QKD transmitter. There, security was proven under the assumption that this specific THA only affects the PM in the transmitter and leaves the other devices untouched. Therefore this result cannot be exported to decoy-state QKD and mdiQKD, where an additional method to modulate the intensity of the prepared signals is required. This is very often achieved via an intensity modulator (IM) inserted in series with the PM. Hence it can happen that partial information about the IM is leaked to Eve, similarly to what happens for the PM. This problem is common to any scheme using devices like PM and IM, such as the decoy-state BB84 protocol~\cite{decoy1,decoy2,decoy3,decoy4,decoy5,decoy6,decoy7,decoy8,decoy9}, bit commitment, oblivious transfer, secure identification~\cite{sbm}, blind quantum computing~\cite{comp} as well as device-independent QKD.

Here we introduce a general formalism to prove the security of most of the optical QKD systems using a PM and an IM in their transmitters that can leak the setting information in an arbitrary manner. As a specific example, we address the optical implementation of the standard decoy-state BB84 QKD protocol with three intensity settings~\cite{decoy1,decoy2,decoy3} due to its extensive use of devices like PM and IM. However, our results can be straightforwardly adapted to any number of settings and to all the protocols mentioned above. Importantly, our approach
is solely based on how the users' devices operate. For a given model of PM and IM, one could readily use our technique to calculate the resulting secret key rate of the system.
This constitutes a fundamental step forward to guaranteeing the security of quantum cryptographic schemes using a PM, an IM or other analogous devices, in presence of information leakage.

To illustrate how our formalism applies to real QKD systems, we investigate a particular form of information leakage, i.e., a THA that is feasible with current technology. In particular, we consider that Eve injects a probe for each phase and intensity setting selected by the legitimate user and the back-reflected light is composed of coherent states of limited intensity.

The paper is organised as follows. In Sec.~\ref{Sec:decoy} we review the main concepts of decoy-state QKD. In Sec.~\ref{Sec: formalism} we present a general formalism to prove its security in the presence of any information leakage from both the PM and the IM. This formalism is then used in
Sec.~\ref{simulations} to study various THA that are feasible with current technology and to evaluate their effect on the system performance.
Finally, Sec.~\ref{discussion} includes a short discussion and Sec.~\ref{concl} concludes the paper with a summary. The paper
also contains Appendixes with calculations that are needed to derive the results in the main text.

\section{Decoy-state quantum key distribution}\label{Sec:decoy}

In decoy-state QKD, Alice prepares mixtures of Fock states with different photon number statistics, selected independently at random for every signal that is sent to Bob. These states can be prepared with practical light sources such as attenuated laser diodes, heralded spontaneous parametric downconversion sources and other practical single-photon sources. They can be formally described as:
\begin{equation}
\rho^\gamma=\sum_{n=0}^\infty p_n^\gamma \ket{n}\bra{n}.
\end{equation}
Here, $p_n^\gamma$ is the photon number statistics, represented by the conditional probability that Alice emits a pulse with $n$ photons when she chooses the intensity setting $\gamma$. The ket $\ket{n}$ denotes an $n$-photon Fock state. If Alice uses a source emitting phase-randomised weak coherent pulses (WCP), the photon number statistics is the Poisson distribution, $p_n^\gamma=e^{-\gamma}\gamma^n/n!$, with $\gamma$ being the mean photon number.

For each intensity setting $\gamma$, there are two quantities which can be directly observed in the experiment: the gain $Q^\gamma=N^\gamma_{\rm click}/N^\gamma$,
where $N^\gamma_{\rm click}$ represents the number of events where Bob observes a click in his measurement device given that Alice prepared the state $\rho^\gamma$, and $N^\gamma$ is the number of signals sent by Alice in the state $\rho^\gamma$,
and the quantum bit error rate (QBER) $E^\gamma=N^\gamma_{\rm error}/N^\gamma_{\rm click}$,
where $N^\gamma_{\rm error}$ denotes the number of errors observed by Bob given that Alice prepared the state $\rho^\gamma$. In the asymptotic limit of large $N^\gamma$
both quantities can be written as a function of the yield $Y_n$ and the error rate $e_n$ of the $n$-photon signals as:
\begin{eqnarray}
\nonumber  Q^{\gamma} &=& \sum_{n=0}^\infty p^{\gamma}_{n}Y_n, \\
E^{\gamma} &=& \frac{1}{Q^{\gamma}}\sum_{n=0}^\infty p^{\gamma}_{n}Y_ne_n,
\label{gain_error}
\end{eqnarray}
for any value of $\gamma$. The unknown parameters in this set of linear equations are $Y_n$ and $e_n$, and they can be estimated by solving Eq.~(\ref{gain_error}).

Indeed, whenever Alice uses an infinite number of settings $\gamma$, any finite set of parameters $Y_n$ and $e_n$ can be estimated with arbitrary precision. If Alice and Bob are only interested in the value of $Y_0$, $Y_1$, and $e_1$, as is the case in QKD, it is possible to obtain a tight estimation of these three parameters
with only a few different intensity settings~\cite{decoy_finite}.
A fundamental implicit requirement in the decoy-state analysis is that the variables $Y_n$ and $e_n$ are independent of the intensity setting $\gamma$. That is, the analysis assumes that Eve does not have any information about Alice's intensity setting choice at each given time. If Eve performs a THA against Alice's source, however, this necessary condition might not be longer satisfied and the security analysis of decoy-state QKD needs to be revised. This is done in the next section.

\section{Trojan horse attacks against decoy-state quantum key distribution}\label{Sec: formalism}

In this section we present a general formalism to evaluate the security of decoy-state QKD against any information leakage from both the IM, which is used to generate decoy-states, and the PM employed to encode the bit and the basis information. Below we assume that such information leakage is due to an active Eve who launches a THA against the decoy-state transmitter. Note, however, that our analysis could be applied as well to any passive information leakage scenario.

\begin{figure}[h!]
\includegraphics[width=0.75\columnwidth]{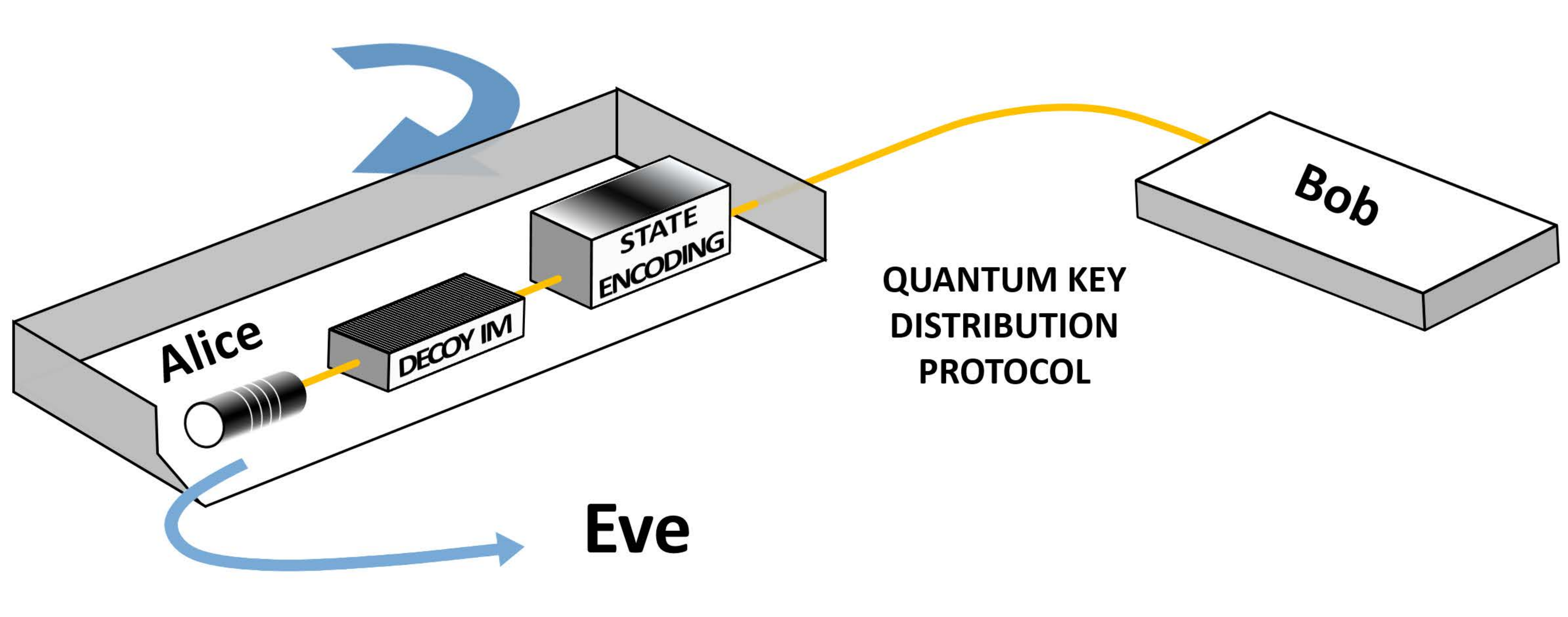}
\centering
\caption{The users Alice and Bob run a QKD protocol with apparatuses that can have leakages (thin arrow in the figure). Any such leaked signal could be captured by Eve and used to steal private information. Eve can even actively shine high-power electromagnetic fields on the system (thick arrow in the figure) to trigger the emission of side-channel signals.}
\label{fig:THA}
\end{figure}

In a THA Eve injects bright light pulses into Alice's device and measures the back-reflected light. This way
she might obtain useful information about Alice's intensity and phase choices for each generated signal.
This situation is illustrated in Fig.~\ref{fig:THA}. As a first consequence, the yields $Y_n$ and the error rates $e_n$ might
now become dependent on the intensity setting $\gamma$, and we will denote them as $Y_n^\gamma$ and $e_n^\gamma$, respectively. The goal of this section is mainly
to evaluate how much can these quantities differ from each other depending on the information leaked to Eve.

\subsection{THA against the IM}\label{subsec:THA against the IM}

Here we focus on the most widely used choice of intensity settings for the standard decoy-state BB84 protocol, where Alice randomly selects one of three possible intensities, denoted as $\gamma_{\rm s}$, $\gamma_{\rm v}$, and $\gamma_{\rm w}$, with probability $p_{\rm s}$, $p_{\rm v}$, and $p_{\rm w}$, respectively. However, our technique can be straightforwardly adapted to cover any number of decoy settings. We will denote as
$\gamma^{i}\in\{\gamma_{\rm s}, \gamma_{\rm v}, \gamma_{\rm w}\}$ the intensity setting selected by Alice in the $i^{th}$ instance of the protocol.

Eve's goal is to learn the value of  $\gamma^{i}$ for all instances $i$. For this, her most general THA can be described as follows. Eve first prepares a probe system ${\rm E_{p}}$, which might be entangled with an ancilla system ${\rm E}$ also in Eve's hands, and sends this system to Alice while she keeps ${\rm E}$ in a quantum memory. The system ${\rm E_{p}}$ may consist of many different pulses, each of them used to probe Alice's intensity setting each given time. Afterwards, Eve performs a joint measurement on all the pulses emitted by Alice together with the systems ${\rm E}$ and the back-reflected light from ${\rm E_{p}}$, which is denoted as ${\rm E_{p}'}$.

Let us consider first the $i^{th}$ $n$-photon pulse emitted by Alice. Later on we will generalise this case to cover all her $n$-photon pulses. For this, let $\rho_{n, \gamma^{i}}$ denote the joint state of Alice's $i^{th}$ $n$-photon pulse and the systems ${\rm E}$ and ${\rm E_{p}'}$~\footnote{For example, if the emission
 of an $n$-photon pulse by Alice is independent of Eve's systems ${\rm E}$ and ${\rm E_{p}'}$, then $\rho_{n, \gamma^{i}}={\hat P}\left(\ket{n}\right)\otimes\rho_{\gamma^{i}}$, where the operator ${\hat P}(\ket{\phi})$ is defined as ${\hat P}(\ket{\phi}):=\ket{\phi}\bra{\phi}$ and $\rho_{\gamma^{i}}$ represents the state of ${\rm E}$ and ${\rm E_{p}'}$. This is the typical situation that one expects in practice.}.
The state of ${\rm E_{p}'}$ may depend on all the intensity choices made by Alice, so does $\rho_{n, \gamma^{i}}$. Now, Eve's task for the $i^{th}$ pulse is to behave as different as possible according to Alice's intensity choice $\gamma^{i}$ given the state $\rho_{n, \gamma^{i}}$. Therefore, we are interested in how well can Eve distinguish the intensity setting $\gamma^{i}_j$ from $\gamma^{i}_k$ and $\gamma^{i}_l$, with $j,k,l\in\{$s,v,w$\}$ and $j\neq k,l$ (note that here $k$ might be equal to $l$). This can be solved using the trace distance argument~\cite{nielsen}, which says that the trace distance between probability distributions arising from any measurement on the states $\rho_{n, \gamma_{j}^{i}}$ and
$q_{nkl}\rho_{n, \gamma_{k}^{i}}+(1-q_{nkl})\rho_{n, \gamma_{l}^{i}}:=\sigma_{n,\gamma_{kl}^i}$ satisfies
\begin{eqnarray}\label{int}
\sum_{\omega\in\Omega}|{\rm Pr}(\omega|\rho_{n, \gamma_{j}^{i}})-{\rm Pr}(\omega|\sigma_{n,\gamma_{kl}^i})|\le 2d\left(\rho_{n, \gamma_{j}^{i}}, \sigma_{n,\gamma_{kl}^i}\right),
\end{eqnarray}
where $\Omega$ is a set of physical events that fulfills $\sum_{\omega\in\Omega}{\rm Pr}(\omega)=1$, $d(\rho,\sigma):={\rm Tr}\left|\rho-\sigma\right|/2$ denotes the trace distance between $\rho$ and $\sigma$, ${\rm Pr}(\omega|\rho)$ is the conditional probability to obtain the event $\omega$ given the state $\rho$, and $q_{nkl}:=p_{k}p_n^{\gamma_k}/(p_{k}p_n^{\gamma_k}+p_{l}p_n^{\gamma_l})$, with $k,l\in\{$s,v,w$\}$, is
the conditional probability to have selected the intensity setting $\gamma_k$ (among only $\gamma_k$ and $\gamma_l$) given that the pulse contains $n$ photons~\footnote{Note that when $k=l$ Eq.~(\ref{int}) implies that $\sum_{\omega\in\Omega}|{\rm Pr}(\omega|\rho_{n, \gamma_{j}^{i}})-{\rm Pr}(\omega|\rho_{n, \gamma_{k}^{i}})|\le 2d(\rho_{n, \gamma_{j}^{i}}, \rho_{n, \gamma_{k}^{i}})$ for all $j,k\in\{$s,v,w$\}$.}.

To prove the security of the decoy-state QKD system, we need to determine Bob's detection rates. This means that we are interested in the set $\Omega=\{{\rm click}, {\rm no\,\,click}\}$, where ``click'' (``no click'') represents a detection (no detection) outcome at Bob's side. That is, Eve must decide which of Alice's pulses will produce (or not produce) a ``click'' at Bob's side
before the quantum part of the protocol finishes. Here, ${\rm Pr}({\rm click}|\rho_{n, \gamma_{j}^{i}})$ is the conditional probability that Bob obtains a ``click'' given $\rho_{n, \gamma_{j}^{i}}$. This probability may depend on the detection pattern observed by Bob in all the previous $i-1$ pulses. By combining Eq.~(\ref{int}) with the fact that ${\rm Pr}({\rm click})+{\rm Pr}({\rm no\,\,click})=1$ we find that
\begin{eqnarray}\label{basis-eq-THA}
|{\rm Pr}({\rm click}|\rho_{n, \gamma_{j}^{i}})-{\rm Pr}({\rm click}|\sigma_{n,\gamma_{kl}^i})|&\le& d\left(\rho_{n, \gamma_{j}^{i}},
\sigma_{n,\gamma_{kl}^i}\right):=D_{n, j, k,l}^{i}.
\end{eqnarray}

Now, in order to relate the conditional probabilities that appear in Eq. (\ref{basis-eq-THA}) with the corresponding actual numbers, we first convert these probabilities
into joint probabilities and then we take the sum over $i\in \{1, 2, \cdots, N\}$, being $N$ the number of trials.
In particular, let ${\rm Pr}({\rm click}, n, \gamma_{j}^{i})$ denote the joint probability that Eve observes the state $\rho_{n, \gamma_{j}^{i}}$ in the instance $i$ and Bob obtains a ``click''. Then, from
Eq.~(\ref{basis-eq-THA})
we obtain that
\begin{eqnarray}\label{basis-eq-THA2}
&&\Bigg|\sum_{i=1}^{N}{\rm Pr}({\rm click}, n, \gamma_{j}^{i})-p_{j}p_n^{\gamma_j}\sum_{i=1}^{N}
\Bigg[q_{nkl}\frac{{\rm Pr}({\rm click}, n, \gamma_{k}^{i})}{p_{k}p_n^{\gamma_k}} +(1-q_{nkl}) \nonumber \\
&&\times\frac{{\rm Pr}({\rm click},n, \gamma_{l}^{i})}{p_{l}p_n^{\gamma_l}}\Bigg]\Bigg| \le p_{j}p_n^{\gamma_j}ND_{n, j, k,l}\,,
\end{eqnarray}
where $D_{n, j, k,l}:=1/N\sum_{i=1}^{N}D_{n, j, k,l}^{i}$.
Importantly, by using Azuma's inequality~\cite{azuma} (see \ref{Explanation of Azuma}), each term on the LHS of Eq.(\ref{basis-eq-THA2})
approaches the actual numbers of the corresponding events except for a probability exponentially small in $N$.
That is, we have that
$\sum_{i=1}^{N}{\rm Pr}({\rm click},n, \gamma_{j}^{i})$ approaches the number of events, $N_{{\rm click}, n, \gamma_{j}}$, within $N$ runs where Alice selects the intensity setting $j$, she emits an $n$-photon state, and Bob obtains a ``click'' in his measurement device. This means that
\begin{eqnarray}\label{basis-eq-THA3}
\left|Y_n^{\gamma_j}-\left[q_{nkl}Y_n^{\gamma_k}+(1-q_{nkl})Y_n^{\gamma_l}\right]\right|&\le& D_{n, j, k,l}\,,
\end{eqnarray}
except for a probability
exponentially small in $N$~\footnote{Note that when $k=l$ Eq.~(\ref{basis-eq-THA3}) implies that $|Y_n^{\gamma_j}-Y_n^{\gamma_k}|\le D_{n, j, k}$ with
$D_{n, j, k}:=1/N\sum_{i=1}^{N}D_{n, j, k}^{i}$ and $D_{n, j, k}^{i}=d(\rho_{n, \gamma_{j}^{i}},\rho_{n, \gamma_{k}^{i}})$. }, where the yields $Y_n^{\gamma_j}$ are defined as
\begin{eqnarray}\label{ups}
Y_n^{\gamma_j}&:=&\frac{N_{{\rm click}, n, \gamma_{j}}}{Np_{j}p_n^{\gamma_j}},
\end{eqnarray}
and similarly for $Y_n^{\gamma_k}$ and $Y_n^{\gamma_l}$. Note that in the special case where there is no information leakage about Alice's intensity choices, we have that  $D_{n, j, k,l}=0$ and, therefore, $Y_n^{\gamma_j}=Y_n^{\gamma_k}=Y_n^{\gamma_l}:=Y_n$, which is the key assumption in the standard decoy-state method (see Sec.~\ref{Sec:decoy}).

The analysis for the error rates $e_n^{\gamma_j}$, with $j\in\{{\rm s, v, w}\}$, is analogous. In particular, here we consider the set $\Omega=\{{\rm click} \wedge {\rm error}, {\rm no\,\,click} \vee ({\rm click} \wedge {\rm no\ error})\}$, where ``click $\wedge$ error'' represents a detection outcome at Bob's side associated with an error,
and ``no click $\vee$ (click $\wedge$ no error)'' denotes a no detection outcome or a detection one associated with no error. Now, taking into account that
${\rm Pr}({\rm click} \wedge {\rm error})+{\rm Pr}[{\rm no\,\,click} \vee ({\rm click} \wedge {\rm no\ error})]=1$, and using a similar analysis as above, we find that
\begin{eqnarray}\label{basis-eq-THA3bb}
\left|Y_n^{\gamma_j}e_n^{\gamma_j}-\left[q_{nkl}Y_n^{\gamma_k}e_n^{\gamma_k}+(1-q_{nkl})Y_n^{\gamma_l}e_n^{\gamma_l}\right]\right|&\le& D_{n, j, k,l}\,,
\end{eqnarray}
where the parameter $D_{n, j, k,l}$ is equal to that given in Eq.~(\ref{basis-eq-THA3})~\footnote{If $k=l$ then Eq.~(\ref{basis-eq-THA3bb}) implies that $|Y_n^{\gamma_j}e_n^{\gamma_j}-Y_n^{\gamma_k}e_n^{\gamma_k}|\le D_{n, j, k}$.}, and $e_n^{\gamma_j}$ is defined as
\begin{eqnarray}
e_n^{\gamma_j}&:=&\frac{N_{{\rm click \wedge error}, n, \gamma_{j}}}{N_{{\rm click}, n, \gamma_{j}}},
\end{eqnarray}
and similarly for $e_n^{\gamma_k}$ and $e_n^{\gamma_l}$. Here, $N_{{\rm click \wedge error}, n, \gamma_{j}}$ represents
the number of events, within $N$ runs, where Alice selects the intensity setting $j$, she emits an $n$-photon state, and Bob obtains a ``click'' associated to an error in his measurement device.

The formalism above is general in the sense that it can be applied to {\it any} THA against Alice's IM. However, to be able to evaluate Eqs.~(\ref{basis-eq-THA3})-(\ref{basis-eq-THA3bb}) one needs to characterise the states $\rho_{n, \gamma_{j}^{i}}$ that are accessible to Eve, and this might be difficult in general. These states are required to calculate the coefficients $D^i_{n, j, k,l}$ and, thus, the parameters $D_{n, j, k,l}$. In the next subsection we show that these parameters can in principle be estimated based solely on the behaviour of the IM.

\subsubsection{Estimation of $D^i_{n, j, k,l}$.}\label{Sec: Trace distance}

In order to
upper bound the value of $D^i_{n, j, k,l}$ based only on how the IM operates, we consider the unitary operator that describes the action of Alice's IM when she selects a certain intensity setting $\gamma^i_j$ for an instance $i$. Importantly,
we assume that this operator characterises the behaviour of the IM on all the optical modes that it supports. That is, in general it acts on Alice's photonic system ${\rm A_{p}}$ ({\it i.e.}, the signal states emitted by her laser), on some additional ancillary system ${\rm A_{a}}$ also in Alice's
hands\footnote{The system ${\rm A_{a}}$ can account for the effect of the loss in Alice's transmitter. That is, we consider that the unitary operator describing her IM
includes as well, together with its intrinsic loss, the effect of any optical attenuator, isolator and filter used by Alice to reduce the energy of the back-reflected light that goes to Eve.}, and on Eve's probe system ${\rm E_{p}}$. Therefore, we will denote it as
${\hat U}_{\rm A_{p}, A_{a}, E_{p}}^{\gamma^i_j}$. 

Let $\ket{\Psi}_{\rm A_{p}, A_{a}, E, E_{p}}$ be the joint state that describes Alice's and Eve's systems before the action of the IM. 
After applying the IM, the state $\ket{\Psi}_{\rm A_{p}, A_{a}, E, E_{p}}$ evolves according to the unitary transformation
${\hat {\mathbbm{1}}}_{\rm E}\otimes{\hat U}_{\rm A_{p}, A_{a}, E_{p}}^{\gamma^i_j}$. Importantly,
in order for the decoy-state method to work, this unitary transformation should produce an output signal with the system 
${\rm A_{p}'}$ (which will be sent to Bob through the quantum channel once the bit and basis information are also encoded) 
prepared in a state that is diagonal in the Fock basis. This is guaranteed if Eve's probing light does not alter the photon distribution of Alice's light source or her phase-randomisation process.
Note here that the physical system corresponding to ${\rm A_{p}'}$ might not be the same as the one for the input system ${\rm A_{p}}$.
This means, in particular, that
\begin{eqnarray}\label{nerga}
{\hat {\mathbbm{1}}}_{\rm E}\otimes{\hat U}_{\rm A_{p}, A_{a}, E_{p}}^{\gamma^i_j}\ket{\Psi}_{\rm A_{p}, A_{a}, E, E_{p}}=\sum_{n}\sqrt{p_{n}^{\gamma^i_j}}
\ket{n^{\gamma^i_j}}_{\rm A_{p}'}\ket{\phi_{n,\Psi}^{\gamma^i_j}}_{\rm A_{a}', E, E_{p}'}\,,
\end{eqnarray}
Here, $p_{n}^{\gamma^i_j}$ denotes the
probability of emitting an $n$-photon pulse in the $i^{th}$ instance of setting ${\gamma_j}$, and $\{\ket{\phi_{n,\Psi}^{\gamma^i_j}}_{\rm A_{a}', E, E_{p}'}\}_{n}$ forms an orthonormal basis, {\it i.e.,} we have that
${}_{\rm A_{a}', E, E_{p}'}\langle\phi_{n',\Psi}^{\gamma^i_j}\ket{\phi_{n,\Psi}^{\gamma^i_j}}_{\rm A_{a}', E, E_{p}'}=\delta_{n'n}$.
Moreover, the physical systems for $\rm A_{a}'$ and $\rm E_{p}'$ might be different from those for $\rm A_{a}$ and $\rm E_{p}$, respectively. 
Also, note that in Eq.~(\ref{nerga}) we have made the general assumption that the photon mode of the $n$-photon state $\ket{n^{\gamma^i_j}}_{\rm A_{p}'}$ might be dependent on the setting ${\gamma^i_j}$.

Now, we focus on those joint states $\ket{n^{\gamma^i_j}}_{\rm A_{p}}\ket{\phi_{n,\Psi}^{\gamma^i_j}}_{\rm A_{a}', E, E_{p}'}$
that contain $n$ photons on Alice's photonic system $\rm A_{\rm p}'$. 
Eve's task is to behave as differently as possible according to the 
intensity setting. We find, therefore, that $D^i_{n, j, k,l}$ can be upper bounded as
\begin{eqnarray}\label{sup-tracebb}
D^i_{n, j, k,l}&\le& {\rm Sup}_{\ket{\Psi}_{\rm A_{p}, A_{a}, E, E_{p}}}\,\,d\Big({\rm Tr}_{\rm A_{a}'}\Big[{\hat P}(\ket{n^{\gamma^i_j}}_{\rm A_{p}'}\ket{\phi_{n,\Psi}^{\gamma^i_j}}_{\rm A_{a}', E, E_{p}'})\Big], \nonumber \\
&&{\rm Tr}_{\rm A_{a}'}\Big[q_{nkl}{\hat P}(\ket{n^{\gamma^i_k}}_{\rm A_{p}'}\ket{\phi_{n,\Psi}^{\gamma^i_{k}}}_{\rm A_{a}', E, E_{p}'}) \nonumber \\
&&+(1-q_{nkl}){\hat P}(\ket{n^{\gamma^i_l}}_{\rm A_{p}'}\ket{\phi_{n,\Psi}^{\gamma^i_{l}}}_{\rm A_{a}', E, E_{p}'})\Big]\Big)\,,
\end{eqnarray}
where the operator ${\hat P}(\ket{\phi}):=\ket{\phi}\bra{\phi}$. This confirms that the description of Alice's IM is enough to guarantee security.

Of course, the formalism above can readily accept any particular assumption on the THA performed by Eve.
For instance, in practical situations it may be over-pessimistic to take the supremum given in Eq.~(\ref{sup-tracebb}) over all possible states $\ket{\Psi}_{\rm A_{p}, A_{a}, E, E_{p}}$. Instead, one might only consider signals of the form $\ket{\Psi}_{\rm A_{p}, A_{a}, E, E_{p}}=\ket{\phi}_{\rm A_{p}}\ket{\varphi}_{\rm A_{a}}\ket{\chi}_{\rm E, E_{p}}$, where
$\ket{\phi}_{\rm A_{p}}$, $\ket{\varphi}_{\rm A_{a}}$ and $\ket{\chi}_{\rm E, E_{p}}$ are pure states of the different systems.
Indeed, this seems to be a natural assumption because Alice's systems ${\rm A}_{\rm p}$ and ${\rm A}_{\rm a}$ are typically independent from each other and also independent from those of Eve. In so doing, Eq.~(\ref{sup-tracebb}) might deliver tighter bounds for $D^i_{n, j, k,l}$.

In general, however, one cannot assume that Eve's state $\ket{\chi}_{\rm E, E_{p}}$ is in a tensor product form. That is, it is not enough to just consider the system ${\rm E_{p}}$ that Eve sends to Alice (together with the back-reflected one) in order to guarantee security. This is so because when the supremum given in Eq.~(\ref{sup-tracebb}) is taken over all joint states $\ket{\chi}_{\rm E, E_{p}}$ it usually results
in a larger trace distance than that obtained when one considers product states. To improve the system performance, Alice might include additional optical elements to force $\ket{\chi}_{\rm E, E_{p}}$ to be of product form. For example, she could perform a phase-randomisation on the system ${\rm E_{p}}$ (see, {\it e.g.},~\cite{ZQL08, ZQL10}). This way all the off-diagonal elements of the state $\ket{\chi}_{\rm E, E_{p}}$ in the Fock basis would vanish, and one could completely disregard system $E$.
Moreover, mathematically, to remove all the off-diagonal elements leads to a significant decrease of the trace distance and, therefore, one expects a significant improvement of the secure key rate, as is confirmed in Sec.~4.3.

\subsection{THA against the PM}
\label{subsec:THA against the PM}

In this section, we review and extend the analysis of the THA against the PM carried out in~\cite{tha3}. The central observation is that the THA allows Eve to partially know Alice's choice of the basis. In other terms, the information leakage is in the form of basis information leaked out to the eavesdropper. This might cause
the density matrices that describe Alice's output states to be {\it basis dependent}. Below, we provide a formalism to prove the security of the BB84 protocol in the presence of the most general THA against the PM.

We will assume that Alice's choice is random, independent of the IM and of the previous preparation instances.
We define the Z basis by the orthogonal vectors $\{ \ket{0}, \ket{1} \}$ and the X basis by $\{ \ket{+}, \ket{-} \}$, where $\ket{\pm}:=(\ket{0}\pm\ket{1})/\sqrt{2}$. We denote as $\ket{\Psi_{\rm Z}^i}_{\rm A_{q}, A_{p}, A_{a}, E, E'_{p}}$ ($\ket{\Psi_{\rm X}^i}_{\rm A_{q}, A_{p}, A_{a}, E, E'_{p}}$) the joint state that describes Alice's
system and Eve's system for the THA given that Alice selected the Z (X) basis.
Here, the superscript $i$ refers to the $i^{th}$ signal generated by Alice, and the system ${\rm A_{q}}$ refers to a virtual qubit that is stored in Alice's lab. Examples of the states $\ket{\Psi_{\rm Z}^i}_{\rm A_{q}, A_{p}, A_{a}, E, E'_{p}}$ and $\ket{\Psi_{\rm X}^i}_{\rm A_{q}, A_{p}, A_{a}, E, E'_{p}}$ are the following
\begin{eqnarray}
\ket{\Psi_{\rm Z}^i}_{\rm A_{q}, A_{p}, A_{a}, E, E'_{p}}=\frac{1}{\sqrt{2}}\left(\ket{0}_{\rm A_{q}} \ket{\Psi^{i}_{0_{\rm Z}}}_{\rm A_{p}, A_{a}, E, E'_{p}}+\ket{1}_{\rm A_{q}} \ket{\Psi^{i}_{1_{\rm Z}}}_{\rm A_{p}, A_{a}, E, E'_{p}} \right), \\
\ket{\Psi_{\rm X}^i}_{\rm A_{q}, A_{p}, A_{a}, E, E'_{p}}=\frac{1}{\sqrt{2}}\left(\ket{+}_{\rm A_{q}} \ket{\Psi^{i}_{0_{\rm X}}}_{\rm A_{p}, A_{a}, E, E'_{p}}+\ket{-}_{\rm A_{q}} \ket{\Psi^{i}_{1_{\rm X}}}_{\rm A_{p}, A_{a}, E, E'_{p}} \right).
\end{eqnarray}
Here, $\ket{\Psi_{j_{\alpha}}^{i}}_{\rm A_{p}, A_{a}, E, E_{p}'}$ (with $j\in\{0,1\}$ and
$\alpha\in\{{\rm Z}, {\rm X}\}$) represents the state of systems ${\rm A_{p}, A_{a}, E}$ and
$\rm E_{p}'$ for Alice's bit value $j$ in her $\alpha$ basis.
We have, therefore, that Alice's state preparation process can be equivalently described as follows. First, she decides which state ($\ket{\Psi_{\rm Z}^i}_{\rm A_{q}, A_{p}, A_{a}, E, E'_{p}}$ or $\ket{\Psi_{\rm X}^i}_{\rm A_{q}, A_{p}, A_{a}, E, E'_{p}}$) she prepares. Afterwards, she measures
the virtual qubit ${\rm A_{q}}$  using the Z or the X basis, depending on the choice of the state. As long as the state preparation is expressed this way, one can
consider any possible purification of the states $\ket{\Psi_{\rm Z}^i}_{\rm A_{q}, A_{p}, A_{a}, E, E'_{p}}$ or $\ket{\Psi_{\rm X}^i}_{\rm A_{q}, A_{p}, A_{a}, E, E'_{p}}$. For instance, one may consider $\ket{\Psi_{\rm X}^i}_{\rm A_{q}, A_{p}, A_{a}, E, E'_{p}}=\frac{e^{{\rm i}\nu}}{\sqrt{2}}\left(\ket{-}_{\rm A_{q}} \ket{\Psi^{i}_{0_{\rm X}}}_{\rm A_{p}, A_{a}, E, E'_{p}}+\ket{+}_{\rm A_{q}} \ket{\Psi^{i}_{1_{\rm X}}}_{\rm A_{p}, A_{a}, E, E'_{p}} \right)$ with $\nu\in[0,2\pi)$ being a global phase. 
Note that we can consider this state because the reduced density operator for systems $\rm A_{p}, A_{a}, E$ and $\rm E'_{p}$ is the same as that of Eq. (13).
The optimal solution is the purification that maximises the key generation rate.

In a security proof, it is essential to determine the phase error rate, which is the parameter needed in the
privacy amplification step of the protocol. The phase error rate is the fictitious bit error rate that
Alice and Bob would have obtained if Alice had measured
the system ${\rm A_{q}}$ with the X basis and Bob had
used the X basis given the preparation of $\ket{\Psi_{\rm Z}^i}_{\rm A_{q}, A_{p}, A_{a}, E, E'_{p}}$.
Intuitively, if the states $\ket{\Psi_{\rm Z}^i}_{\rm A_{q}, A_{p}, A_{a}, E, E'_{p}}$ and $\ket{\Psi_{\rm X}^i}_{\rm A_{q}, A_{p}, A_{a}, E, E'_{p}}$ are close
enough to each other, then the
phase error rate should be close to the X basis error rate which is obtained in the actual experiment. Below, we make this argument more rigorous
by using the analysis presented in~\cite{LP06}. For this, we will assume that
the basis choice is done in a coherent manner, i.e., Alice first prepares the joint system
\begin{eqnarray}
\ket{\Psi^{i}}_{\rm A_c, A_{q}, A_{p}, A_{a}, E, E'_{p}}=\frac{1}{\sqrt{2}}\left(\ket{0}_{\rm A_c}\ket{\Psi^{i}_{\rm Z}}_{\rm A_{q}, A_{p}, A_{a}, E, E'_{p}}+\ket{1}_{\rm A_c}\ket{\Psi^{i}_{\rm X}}_{\rm A_{q}, A_{p}, A_{a}, E, E'_{p}}\right),\nonumber\\
\label{Q-coin}
\end{eqnarray}
where the system ${\rm A_c}$ is the so-called ``quantum coin''~\cite{gllp}.
Importantly, the phase error rate is related to the X basis measurement
on the quantum coin.
To derive the formula for the estimation of the phase error rate, we consider the following fictitious protocol.
In particular,
for the $i^{th}$ trial of the protocol, Alice and Eve prepare their systems in the state $\ket{\Psi^{i}}_{\rm A_{c}, A_q, A_{p}, A_{a}, E, E'_{p}}$, Alice keeps systems ${\rm A_{a}}$, ${\rm A_{q}}$
and ${\rm A_{c}}$ in her hands, and
sends system $\rm A_{p}$ to Bob. At the reception side,
Bob receives some optical systems after Eve's intervention, and he performs the X basis measurement.
In addition, Alice
performs the X basis measurement on the system ${\rm A_{q}}$.
Then, Alice randomly chooses between the Z or the X basis with equal probability to measure her quantum coin ${\rm A_c}$. Here, note that, from Eq.~(\ref{Q-coin}), when
Alice chooses the Z basis to measure the coin and the result is ``$0$'' (``$1$''),
this is equivalent to Alice and Eve directly preparing the state $\ket{\Psi_{\rm Z}^i}_{\rm A_{q}, A_{p}, A_{a}, E, E'_{p}}$ ($\ket{\Psi_{\rm X}^i}_{\rm A_{q}, A_{p}, A_{a}, E, E'_{p}})$. Next, 
we apply the Bloch sphere bound~\cite{Blochspherebound}
for probability distributions to those instances where Bob obtained a
click event. In particular, we first apply this bound separately to the events with the X basis error and to those with no X basis error. We obtain the following two inequalities
\begin{eqnarray}\label{martemp}
1-2{\rm Pr}^{i}({\rm X_{Ac}}=-|{\rm X}-{\rm Error})\nonumber\\
\quad \quad\le2\sqrt{{\rm Pr}^{i}({\rm Z_{Ac}}=1|{\rm X}-{\rm Error})(1-{\rm Pr}^{i}({\rm Z_{Ac}}=1|{\rm X}-{\rm Error}))},\\
1-2{\rm Pr}^{i}({\rm X_{Ac}}=-|{\rm No}\ {\rm X}-{\rm Error})\nonumber\\
\quad \quad\le2\sqrt{{\rm Pr}^{i}({\rm Z_{Ac}}=1|{\rm No}\ {\rm X}-{\rm Error})(1-{\rm Pr}^{i}({\rm Z_{Ac}}=1|{\rm No}\ {\rm X}-{\rm Error}))}\,.\nonumber\\
\label{martemp2}
\end{eqnarray}
Here, ${\rm Pr}^{i}({\rm X_{Ac}}=-|{\rm X}-{\rm Error})$ is the conditional probability of observing the outcome ``$-$'' when performing the X basis measurement on the quantum coin given that there is a X basis error; ${\rm Pr}^{i}({\rm Z_{Ac}}=1|{\rm X}-{\rm Error})$ is the conditional probability of observing
the outcome ``$1$'' when performing the Z basis measurement on the quantum coin given that there is a X basis error; and the other probabilities are defined similarly. Next, we multiply both inequalities by the term
${\rm Pr}^{i}({\rm click})$, which is
the probability that Bob obtains a ``click'' in his measurement apparatus, and after combining Eqs.~(\ref{martemp})-(\ref{martemp2}) we obtain~\cite{LP06}
\begin{eqnarray}
{\rm Pr}^{i}({\rm click})-2 {\rm Pr}^{i}({\rm X}_{\rm A_c}=-) \le 2\sqrt{{\rm Pr}^{i}({\rm X}, {\rm X}-{\rm Error}){\rm Pr}^{i}({\rm Z}, {\rm X}-{\rm Error})} \nonumber \\
\quad \quad \quad\quad \quad \quad\quad \quad+2\sqrt{{\rm Pr}^{i}({\rm X}, {\rm No}\ {\rm X}-{\rm Error}){\rm Pr}^{i}({\rm Z}, {\rm No}\ {\rm X}-{\rm Error})},
\label{Bloch sphere}
\end{eqnarray}
\noindent where ${\rm Pr}^{i}({\rm X}_{\rm A_c}=-)$ is the probability that the measurement result on the quantum coin is ``$-$'',
${\rm Pr}^{i}({\rm X}, {\rm X}-{\rm Error})$ is the joint probability of selecting the Z basis to measure the quantum coin and obtaining the result ``$1$'' (which implies the preparation of the state
$\ket{\Psi_{\rm X}^i}_{\rm A_{q}, A_{p}, A_{a}, E, E'_{p}}$), and observing a bit error in Alice's and Bob's X basis measurement. The probability
${\rm Pr}^{i}({\rm Z}, {\rm X}-{\rm Error})$ is the fictitious joint probability of selecting the Z basis to measure the quantum coin,
and obtaining the result ``$0$'' (which implies the preparation of the state
$\ket{\Psi_{\rm Z}^i}_{\rm A_{q}, A_{p}, A_{a}, E, E'_{p}}$), and observing a bit error in Alice's and Bob's X basis measurement.
Actually, this last probability is the phase error rate. The probabilities ${\rm Pr}^{i}({\rm X}, {\rm No}\ {\rm X}-{\rm Error})$ and
${\rm Pr}^{i}({\rm Z}, {\rm No}\ {\rm X}-{\rm Error})$
are defined in a similar way (see~\cite{LP06} for further details).
Note that in order to obtain Eq.~(\ref{Bloch sphere}) from Eqs.~(\ref{martemp})-(\ref{martemp2}) we have used the fact that ${\rm Pr}^i({\rm X_{Ac}=-}, {\rm click})\le {\rm Pr}^i({\rm X_{Ac}=-})$, where ${\rm Pr}^i({\rm X_{Ac}=-}, {\rm click})$ represents the joint probability that the measurement result on the quantum coin is ``$-$'' and Bob obtains a ``click'' event with his measurement.
Importantly,
the probability
${\rm Pr}^{i}({\rm X}_{\rm A_c}=-)$ characterises how close are the states $\ket{\Psi_{\rm Z}^i}_{\rm A_{q}, A_{p}, A_{a}, E, E'_{p}}$ and $\ket{\Psi_{\rm X}^i}_{\rm A_{q}, A_{p}, A_{a}, E, E'_{p}}$. Specifically, by choosing an appropriate global phase for $\ket{\Psi_{\rm X}^i}_{\rm A_{q}, A_{p}, A_{a}, E, E'_{p}}$,
from Eq.~(\ref{Q-coin}) we have that
\begin{eqnarray}
{\rm Pr}^{i}({\rm X}_{\rm A_c}=-)=\frac{1}{2}\left(1-\left|{}_{\rm A_{q}, A_{p}, A_{a}, E, E'_{p}}\bra{\Psi_{\rm Z}^i}\Psi_{\rm X}^i\rangle_{\rm A_{q}, A_{p}, A_{a}, E, E'_{p}}\right|\right)\,.
\end{eqnarray}
The term $|{}_{\rm A_{q}, A_{p}, A_{a}, E, E'_{p}}\bra{\Psi_{\rm Z}^i}{\Psi_{\rm X}^i}\rangle_{\rm A_{q}, A_{p}, A_{a}, E, E'_{p}}|$ can be
upper-bounded by the fidelity between the Z basis state and the X basis state. This means that
Eq.~(\ref{Bloch sphere}) gives us the phase error probability
taking into account the ``closeness'' between the two basis states.
To relate the probabilities with the actual number of the corresponding events, we first
use the concavity of the square root function and we take the sum
over $i\in\{1, 2, \cdots, N\}$, with $N$ being
the number of pulses sent in the fictitious protocol. In so doing, we find that
\begin{eqnarray}
&&\sum_{i=1}^{N}{\rm Pr}^{i}({\rm click}) -2\sum_{i=1}^{N} {\rm Pr}^{i}({\rm X}_{\rm A_c}=-)\nonumber\\
&\le& 2\sqrt{\left[\sum_{i=1}^{N}{\rm Pr}^{i}({\rm X}, {\rm X}-{\rm Error})\right] \left[\sum_{i=1}^{N}{\rm Pr}^{i}({\rm Z}, {\rm X}-{\rm Error})\right]} \nonumber\\
&+&2\sqrt{\left[\sum_{i=1}^{N}{\rm Pr}^{i}({\rm X}, {\rm No}\ {\rm X}-{\rm Error})\right] \left[\sum_{i=1}^{N}{\rm Pr}^{i}({\rm Z}, {\rm No}\ {\rm X}-{\rm Error})\right]}\,.
\label{Bloch sphere2}
\end{eqnarray}
\noindent Next, we apply Azuma's inequality~\cite{azuma} (see \ref{Explanation of Azuma}). We obtain, therefore, that except for a probability exponentially small in $N$
each sum
of the probability distributions approaches the actual number of the
corresponding events in $N$ trials. That is,
\begin{eqnarray}
1-2\frac{N_{{\rm X}_{\rm A_c}=-}}{N_{{\rm click}}}&\le& 2\sqrt{\frac{N_{{\rm X}, {\rm X}-{\rm Error}}}{N_{{\rm click}}}\frac{N_{{\rm Z}, {\rm X}-{\rm Error}}}{N_{{\rm click}}}}\nonumber\\
&+&2\sqrt{\frac{N_{{\rm X}, {\rm No}\ {\rm X}-{\rm Error}}}{N_{{\rm click}}}\frac{N_{{\rm Z}, {\rm No}\ {\rm X}-{\rm Error}}}{N_{{\rm click}}}}\,,
\label{Bloch sphere3}
\end{eqnarray}
where $N_g$ denotes the number of instances associated to the event $g$.
Importantly, here $N_{{\rm Z}, {\rm X}-{\rm Error}}/N_{\rm click}$ is
related to the phase error rate, that is, the rate of choosing the $Z$ basis and having the phase error, and
$N_{{\rm X}, {\rm X}-{\rm Error}}/N_{\rm click}$ is the observed ratio of choosing the $X$ basis and having a bit error.
As for $N_{{\rm X}_{\rm A_c}=-}$, we have that except for a probability exponentially small in $N$ the following inequality
is satisfied. 
\begin{eqnarray}
N_{{\rm X}_{\rm A_c}=-}&\le&\sum_{i=1}^{N}\frac{1-|{}_{\rm A_{q}, A_{p}, A_{a}, E, E'_{p}} \bra{\Psi^i_{\rm Z}}\Psi^i_{\rm X}\rangle_{\rm A_{q}, A_{p}, A_{a}, E, E'_{p}}|}{2} \nonumber \\
&\le&
\frac{N}{2}\left[1-\min_{i}|{}_{\rm A_{q}, A_{p}, A_{a}, E, E'_{p}} \bra{\Psi^i_{\rm Z}}\Psi^i_{\rm X}\rangle_{\rm A_{q}, A_{p}, A_{a}, E, E'_{p}}|\right]\,.
\end{eqnarray}
This is so because we can directly calculate the probability ${\rm Pr}^{i}({\rm X}_{\rm A_c}=-)$ from Eq. (\ref{Q-coin}).
Therefore, if Alice and Bob know the minimum overlap between the states $\ket{\Psi^i_{\rm X}}_{\rm A_{q}, A_{p}, A_{a}, E, E'_{p}}$ and $\ket{\Psi^i_{\rm Z}}_{\rm  A_{q}, A_{p}, A_{a}, E, E'_{p}}$
they can estimate the value of the phase error rate even if Eve performs the
most general THA against the PM. The estimation of such overlap, however, might be difficult in general
as one would need to know
Eve's ancilla state. To overcome this problem, we proceed like in the previous section
and we reformulate the formalism above based only on how the PM operates.

For this, note that $\ket{\Psi^i_{\rm Z}}_{\rm A_{q}, A_{p}, A_{a}, E, E'_{p}}$ and
$\ket{\Psi^i_{\rm X}}_{\rm  A_{q}, A_{p}, A_{a}, E, E'_{p}}$ can be expressed as
\begin{eqnarray}
\ket{\Psi^{i}_{\zeta}}_{\rm A_{q}, A_{p}, A_{a}, E, E'_{p}}:={\hat {\mathbbm{1}}}_{\rm A_{q}, E}\otimes{\hat U}_{\rm A_{p}, A_{a}, E_{p}}^{\zeta, i} \ket{\Psi}_{\rm A_{q}, A_{p}, A_{a}, E, E_{p}} \,,
\label{Unitary for THA PM}
\end{eqnarray}
where $\zeta\in\{{\rm X, Z}\}$, and ${\hat U}_{\rm A_{p}, A_{a}, E_{p}}^{\zeta,i}$\footnote{Similar to the IM, in general, the PM and other devices, may be correlated in their operations. In this case, this unitary transformation could depend on all the previous intensity choices that Alice has already made.} is the $i^{th}$ unitary transformation associated to the PM. It supports
Alice's photonic system ${\rm A_{p}}$ and her ancilla ${\rm A_{a}}$, and Eve's ancilla ${\rm E}$ together with her probe system ${\rm E_{p}}$. With this
unitary transformation, the overlap between $\ket{\Psi^i_{\rm Z}}_{\rm A_{q}, A_{p}, A_{a}, E, E'_{p}}$ and
$\ket{\Psi^i_{\rm X}}_{\rm A_{q}, A_{p}, A_{a}, E, E'_{p}}$ for the $i^{th}$ instance
can be lower-bounded as
\begin{eqnarray}
{\rm Inf}_{\ket{\Psi}_{\rm A_{q}, A_{p}, A_{a} E, E_{p}}}
\left| {}_{\rm  A_{q}, A_{p}, A_{a}, E, E_{p}}\bra{\Psi} {\hat U}_{\rm A_{p}, A_{a}, E_{p}}^{{\rm Z}, i\dagger}{\hat U}_{\rm A_{p}, A_{a}, E_{p}}^{{\rm X}, i}  \ket{\Psi}_{\rm A_{q}, A_{p}, A_{a}, E, E_{p}}  \right|\,,
\end{eqnarray}
which is independent of the state.
Note that here we have used the infimum because the unitary operator could support a mode in a Hilbert space containing an arbitrary number of photons.
Therefore, Eq. (\ref{Bloch sphere3}) can be written as 
\begin{eqnarray}
&&1-\frac{N}{N_{\rm click}}\Bigg(1-\min_{i}{\rm Inf}_{\ket{\Psi}_{\rm A_{q}, A_{p}, A_{a} E, E_{p}}}
\Bigg| {}_{\rm A_{q}, A_{p}, A_{a}, E, E_{p}}\bra{\Psi}{\hat U}_{\rm A_{p}, A_{a}, E_{p}}^{{\rm Z}, i\dagger}{\hat U}_{\rm A_{p}, A_{a}, E_{p}}^{{\rm X}, i}  \nonumber \\
&\times&\ket{\Psi}_{\rm A_{q}, A_{p}, A_{a}, E, E_{p}}  \Bigg|\Bigg)
\le 2\sqrt{\frac{N_{{\rm X}, {\rm X}-{\rm Error}}}{N_{\rm click}}\frac{N_{{\rm Z}, {\rm X}-{\rm Error}}}{N_{\rm click}}} \nonumber \\
&+&2\sqrt{\frac{N_{{\rm X}, {\rm No}\ {\rm X}-{\rm Error}}}{N_{\rm click}}\frac{N_{{\rm Z}, {\rm No}\ {\rm X}-{\rm Error}}}{N_{\rm click}}}\,.
\end{eqnarray}
Finally, we use
\begin{eqnarray}
\delta_{{\rm X}-{\rm Error}|{\rm X}}&:=&\frac{N_{{\rm X}, {\rm X}-{\rm Error}}}{N_{{\rm X}}},
\quad \quad \delta_{{\rm No}\ {\rm X}-{\rm Error}|{\rm X}}:=\frac{N_{{\rm X}, {\rm No}\ {\rm X}-{\rm Error}}}{N_{\rm X}},
\nonumber \\
\delta_{{\rm X}-{\rm Error|{\rm Z}}}&:=&\frac{N_{{\rm Z}, {\rm X}-{\rm Error}}}{N_{\rm Z}}, \quad \quad
\delta_{{\rm No}\ {\rm X}-{\rm Error|{\rm Z}}}:=\frac{N_{{\rm Z}, {\rm No}\ {\rm X}-{\rm Error}}}{N_{\rm Z}},
\end{eqnarray}
where $N_{\rm Z}$ ($N_{\rm X}$) is the number of events where Alice's Z-basis measurement outcome on the quantum coin is ``$0$''
(``$1$''). That is, Alice prepares the Z-basis (X-basis) state and Bob detects signals in the Z basis (X basis) in the actual protocol
(recall that the virtual protocol concentrates only on the basis matched events). Then, by taking into account that
$\sqrt{x(1-x)}\le1/2$ for $0\le x\le1$, we obtain the following modified inequality
\begin{eqnarray}\label{auto}
&&1-\frac{N}{N_{\rm click}}\Bigg(1-\min_{i}{\rm Inf}_{\ket{\Psi}_{\rm A_{q}, A_{p}, A_{a} E, E_{p}}}
\Bigg| {}_{\rm A_{q}, A_{p}, A_{a}, E, E_{p}}\bra{\Psi}{\hat U}_{\rm A_{p}, A_{a}, E_{p}}^{{\rm Z}, i\dagger}{\hat U}_{\rm A_{p}, A_{a}, E_{p}}^{{\rm X},i}  \nonumber \\
&&
\ket{\Psi}_{\rm A_{q}, A_{p}, A_{a}, E, E_{p}}  \Bigg|\Bigg)
\le \sqrt{\frac{N_{\rm X} \delta_{{\rm X}-{\rm Error}|{\rm X}}}{N_{\rm click}}\frac{N_{\rm Z}\delta_{{\rm X}-{\rm Error}|{\rm Z}}}{N_{\rm click}}} \nonumber \\
&+&\sqrt{\frac{N_{\rm X} (1-\delta_{{\rm X}-{\rm Error}|{\rm X}})}{N_{\rm click}}\frac{N_{\rm Z}(1-\delta_{{\rm X}-{\rm Error}|{\rm Z}})}{N_{\rm click}}}.
\end{eqnarray}
Remember that $N_{\rm click}$ represents the number of detected events by Bob in the {\it actual} protocol since
the quantum coins have been measured along the Z basis, which corresponds to the case in the actual protocol. Therefore, we have that the RHS of this equation is consistent with the results presented in~\cite{LP06}.

Like in the previous section, note that the formalism above
can readily accept any assumption on the THA. For example, if one considers a
specific THA against the PM where Alice and Bob know the fidelity $F_{\rm X,Z}$ between the
two density matrices describing the output states for the X and Z bases, we have that
\begin{eqnarray}
1-\frac{N}{N_{\rm click}}\left(1-F_{\rm X,Z}\right)&\le& \sqrt{\frac{N_{\rm X} \delta_{{\rm X}-{\rm Error}|{\rm X}}}{N_{\rm click}}\frac{N_{\rm Z}\delta_{{\rm X}-{\rm Error}|{\rm Z}}}{N_{\rm click}}} \nonumber \\
&+&\sqrt{\frac{N_{\rm X} (1-\delta_{{\rm X}-{\rm Error}|{\rm X}})}{N_{\rm click}}\frac{N_{\rm Z}(1-\delta_{{\rm X}-{\rm Error}|{\rm Z}})}{N_{\rm click}}},
\end{eqnarray}
which is essentially the result obtained in~\cite{tha3}. This means in particular that with the estimation of the fidelity given for an explicit THA, as the one considered in the next section, one can readily obtain the phase error rate and therefore the secure key rate of a QKD system endowed with a leaky PM.

Until now we have discussed the scenario where the THA against the IM and the PM acts independently on these two devices.
However, in general, the IM and the PM might present correlations which could be exploited by Eve in a joint THA.
More specifically, the leaked information might be dependent on both the intensity setting and the bit and basis choices.
This situation is addressed in~\ref{Correlation}, where we discuss how to adapt the formalism above to also cover this case. 

\section{Simulation of the key generation rate}
\label{simulations}

In order to apply the theoretical description to a practical case, we treat the THA as a particular form of information leakage, actively caused by the eavesdropper. We draw a realistic worst-case scenario following the line of Ref.~\cite{tha3}, where a THA targeting the PM placed in Alice's box was studied. Here, we review this argument and employ it to any other device that is actively modulated in the transmitting unit, in particular to the IM that is commonly employed to run a decoy-state protocol. We assume that Eve uses a continuous-wave (CW) high-power laser to probe a QKD transmitter. The suitability of a CW laser for the THA is due to a twofold reason. Firstly, it is less destructive than a pulsed laser~\cite{wood}, so it is less easily detectable by Alice and Bob. Secondly, a CW laser is not less efficient than a pulsed laser in probing devices that are modulated according to a non-return-to-zero (NRZ) logic, and assuming NRZ modulation for the transmitter's devices is a conservative choice~\cite{tha3}. Also, it is apparent that the THA is enhanced if the power of Eve's laser is as large as possible, because this maximises the amount of back-reflected light for any fixed reflectivity of the transmitting unit. Therefore we can think that Eve's laser is operated well above threshold.

A consequence of these preliminary considerations is that it is not too restrictive in practice to consider a THA performed with a CW laser operated well above threshold. In turn, such a laser emits light in a state that is closely approximated by a single-mode coherent state~\cite{loudon}. We will therefore assume in this section that Eve uses high-intensity single-mode coherent states to perform the THA. Formally, we write the input coherent state as $|\beta' e^{i\theta'}\rangle$, where $\beta'$ is a real number representing the amplitude of the input light and $\theta'$ is an arbitrary phase that can be set equal to zero without loss of generality. Notice that even if Eve's laser is CW, it still makes sense to use the expression ``light pulse'' for Eve's light, as a light pulse is temporally defined by Eve to match the modulation period of the transmitter's devices.
When a coherent state of light enters the QKD transmitter, it undergoes transformations that are linear and cannot change its photon statistics. So the light back-reflected to Eve will still be in a coherent state, which we indicate as:
\begin{equation}\label{coh-state}
  |\beta_{\gamma_j} e^{i\theta_{\gamma_j}}\rangle.
\end{equation}
In this case, the real numbers $\beta_{\gamma_j}$ and $\theta_{\gamma_j}$ are amplitude and phase, respectively, of the light back-reflected to Eve, which can depend on the intensity setting of the transmitter, $\gamma_j$. Notice though that they are assumed not to depend on the particular instance $i$ of the preparation. Moreover, in writing Eq.~(\ref{coh-state}), we assume that there is no entanglement between Alice's system ${\rm A_p}$ and Eve's probe system ${\rm E'_p}$. Therefore we term ``individual'' this particular class of THA.

In the next sections, we will simulate the secure key rate of a typical decoy-state QKD system against the individual THA, in three different cases of practical interest, with the aim to provide security guidelines of immediate use in QKD experiments.
The three cases correspond to different assumptions about the state in Eq.~(\ref{coh-state}), which will be described in detail in the next Secs.~\ref{subsec:indTHA1}, \ref{subsec:indTHA2} and \ref{subsec:indTHA3}. These cases will be also schematically summarised in Fig.~\ref{fig:models}, at the end of this section. However, Fig.~\ref{fig:models} could even be used as an introductory scheme to our models instead, as it conveniently displays the assumptions underlying the simulations.

To draw the simulations, the main ingredient is the characterisation of the transmitters' modulators, which, as discussed in the previous section, leads to upper bound the trace distance between the different settings of the modulators in the presence of leaked information, as described by Eq.~(\ref{sup-tracebb}) (see also \ref{app_est} and \ref{appen_parD}). In practice, this often translates into defining the modes transmitted by the modulators and their attenuation coefficients. Then, a specific protocol can be considered and its secure key rate estimated. In the simulations, we will consider the following lower bound to the asymptotic secure key rate of the decoy-state BB84 protocol~\cite{qkd1}:
\begin{equation}\label{rate-eq}
  K \geq \max_{\Gamma_{\rm A}} ~\min_{\Gamma_{\rm E}}~q \{ p_0^{\gamma_{\rm s}} Y_{0{\rm L}}^{\gamma_{\rm s}} + p_1^{\gamma_{\rm s}} Y_{1{\rm L}}^{\gamma_{\rm s}} [ 1-h(e_{1{\rm U}}^{\gamma_{\rm s}}) ] -f(E^{\gamma_{\rm s}}) Q^{\gamma_{\rm s}} h(E^{\gamma_{\rm s}})\},
\end{equation}
where $\Gamma_{\rm A}$ and $\Gamma_{\rm E}$ are the spaces of the parameters controlled by Alice and by Eve, respectively. In the simulation, we will use $\Gamma_{\rm A}=\{ \gamma_{\rm s}, \gamma_{\rm v}\}$ and $\Gamma_{\rm E}= \{\theta_{\gamma_j}\}$, and assume without loss of generality that $\gamma_{\rm s}\geq\gamma_{\rm v}\geq\gamma_{\rm w}$ and $\theta_{\gamma_s}=0$, $\theta_{\gamma_v}\in[0,2\pi]$, $\theta_{\gamma_w}\in[\theta_{\gamma_v},2\pi]$.
Here, as for $\gamma_{\rm w}$ and $\beta_{\gamma_{j}}$, we will fix them to particular constant values in the simulation.
In Eq.~(\ref{rate-eq}), the key is distilled only from the signal states; $q$ is the efficiency of the protocol; $p_0^{\gamma_{\rm s}}=e^{-{\gamma_{\rm s}}}$ and $p_1^{\gamma_{\rm s}}=\gamma_{\rm s} e^{-{\gamma_{\rm s}}}$; $Y_{0{\rm L}}^{\gamma_{\rm s}}$ and $Y_{1{\rm L}}^{\gamma_{\rm s}}$ ($e_{1{\rm U}}^{\gamma_{\rm s}}$) are lower (upper) bounds for $Y_{0}^{\gamma_{\rm s}}$ and $Y_{1}^{\gamma_{\rm s}}$, respectively, ($e_{1}^{\gamma_{\rm s}}$) is defined in Sec.~\ref{Sec: formalism}; $f(E^{\gamma_{\rm s}})$ is the efficiency of the error correction protocol; $h(x)=-x\log_2(x)-(1-x)\log_2(1-x)$ is the binary Shannon entropy function. All the parameters used in the simulation are listed in Table~\ref{table1} and the associated physical model for the quantum transmission is described in~\ref{toolbox}. The calculation of $K$ passes through the estimation of $Y_{0{\rm L}}^{\gamma_{\rm s}}$, $Y_{1{\rm L}}^{\gamma_{\rm s}}$ and $e_{1{\rm U}}^{\gamma_{\rm s}}$, which is performed by numerical constrained optimisation as explained in~\ref{app_est}.
\begin{table}[htb]
  \centering
  \begin{tabular}{lccccccc}
    \hline\hline
  $q$ & \quad $e_{\rm d}$ \quad & \quad $p_{\rm d}$ \quad & \quad $\eta_{\rm B}$ \quad & \ \quad $\eta_{\rm det}$ \ \quad & \ \quad $\alpha$ \ \quad & \ \quad $\gamma_{\rm w}$ \ \quad & \ \quad $f(E^{\gamma_{\rm s}})$ \ \quad \\
  \hline
  1 & \ \quad\quad 0.01 \ \quad\quad & \quad $5\times 10^{-6}$ \quad  & \quad 0.5 \quad  & \quad 0.25  \quad  & \quad 0.2  \quad  & \quad $5\times 10^{-4}$  \quad  & \quad 1.2  \quad  \\
 \hline\hline
\end{tabular}
\caption{Experimental parameters used in the simulation of the secure key rate. The associated physical model is explained in~\ref{toolbox}. The values reported in the table are commonly met in a fibre-based QKD setup, see e.g.~\cite{LPD+13}. The intensity parameters $\gamma_{\rm s}$ and $\gamma_{\rm v}$ are not displayed in the table as they are optimised numerically at every distance. The parameter $\gamma_{\rm w}$ is set equal to a constant value to reduce the parameter space of the simulation. Its effect on the key rate is marginal.}
\label{table1}
\end{table}

\subsection{Individual THA - Case~1}
\label{subsec:indTHA1}

\begin{figure}[h!]
\includegraphics[width=0.65\columnwidth]{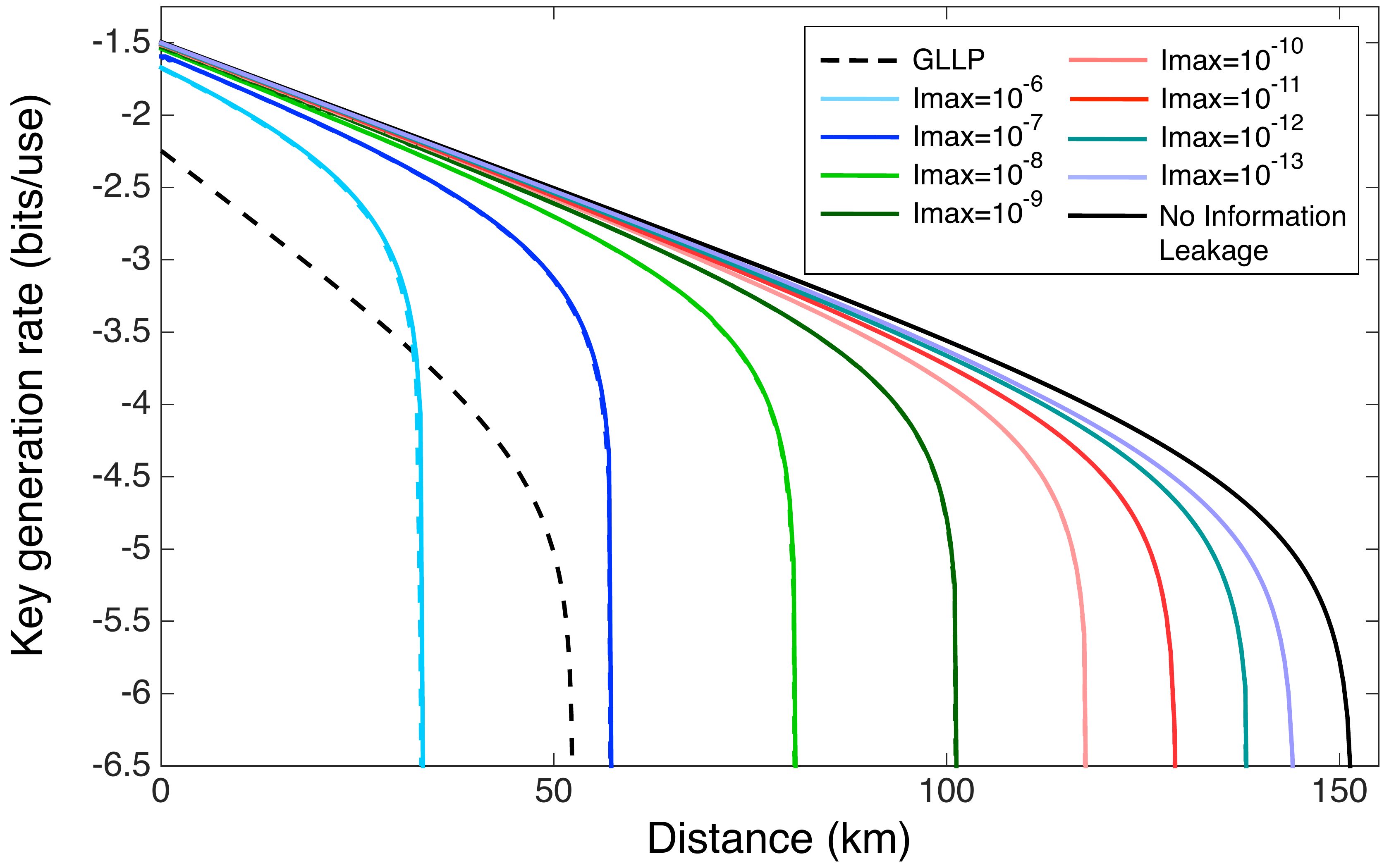}
\centering
\caption{Secure key rate versus distance in presence of a THA targeting the modulating devices of a QKD transmitter. Each colour corresponds to a different value of the intensity of the leaked light, $I_{\rm max}$. The depicted key rate is for the worst-case of a single value of $I_{\rm max}$ bounding all the intensity settings in the transmitter, see Eq.~(\ref{set1}) in the main text.  The solid lines are for a leakage due only to the IM, while the dashed lines, visible for $I_{\rm max}$ equal to $10^{-6}$, $10^{-7}$ and $10^{-8}$, are for the total leakage coming from IM and PM simultaneously. For every distance, the key rate is minimised over the angles $\theta_{\gamma_j}$, controlled by Eve, and maximised over the amplitudes ${\rm \gamma_s}$ and ${\rm \gamma_v}$, controlled by Alice. All the parameters used in the simulation are listed in Table~\ref{table1}.}
\label{fig:case1}
\end{figure}

As mentioned in Sec.~\ref{Sec: formalism}, Eve's goal in a THA is to maximise the difference between the states leaked out of the transmitter. Because these are represented by the coherent state in Eq.~(\ref{coh-state}), Eve's task is simpler when the intensity of the relevant states is larger, as this makes the states more orthogonal. Therefore, the first scenario we consider is one in which we over-estimate the intensity of the leaked states so to draw a consistent worst-case scenario for the individual THA. Suppose that the users characterise their apparatus and find that the intensity of the leaked signals is always upper bounded by a certain value $I_{\rm max}$. This could be the result of an experiment aimed at characterising the worst-case reflectivity of the transmitter as a whole, without specifically addressing the individual devices inside the transmitting unit. Because in the estimation of the secure key rate, Eq.~(\ref{rate-eq}), we assume that the parameters $\theta_{\gamma_j}$ are entirely controlled by Eve, it is conservative to set the intensities of the states leaked out from the transmitter as follows:
\begin{equation}\label{set1}
  \beta_{\rm \gamma_s}^2 = \beta_{\rm \gamma_v}^2 = \beta_{\rm \gamma_w}^2 = \beta^2 = I_{\rm max}.
\end{equation}
The detailed calculation of the trace distance terms $D_{n,j,k,l}$ for the leaked states under the settings of Eq.~(\ref{set1}) is given in~\ref{subsec:appen_parD_Case1}. Then, the key rate in Eq.~(\ref{rate-eq}) is numerically simulated and the result is plotted in Fig.~\ref{fig:case1} as a function of the distance between the users.
The colours correspond to different values of the parameter $I_{\rm max}$. The black solid line represents the ideal case of no information leakage. When the information leakage intensity is lower than $10^{-6}$ photons/pulse, it is always possible to distill a secure quantum key, even in presence of the THA.
When $I_{\rm max}=10^{-6}$, the key rate distilled from our security proof remains positive up to distances of about 30~km. This can be compared with implementation without decoy states, where a single unmodulated intensity is used. In this case, the so-called GLLP security proof~\cite{gllp} applies, and the corresponding key rate is depicted in Fig.~\ref{fig:case1} as a dashed black line.
When $I_{\rm max}$ is smaller than $10^{-12}$, the key rate in presence of a THA approaches closely that of a perfectly shielded system over short and medium-range distances, whereas it deviates from ideal over longer distances. In this latter case, a non-negligible amount of additional privacy amplification is required to protect the system against the THA.
In the same figure, we also include dashed coloured lines to represent the secure key rate in presence of a THA that targets simultaneously the IM and the PM enclosed in a QKD transmitter. For that, we conservatively assumed that Eve gets the same amount of back-reflected light, $I_{\rm max}$, from the IM and the PM separately, so to maximise her information gain about each modulator. As it is apparent from Fig.~\ref{fig:case1}, the lines corresponding to this case are almost perfectly overlapping with the lines corresponding to having only the IM attacked by the THA. This suggests that protecting the IM of a decoy-state QKD transmitter against the THA is more challenging than protecting the PM alone. In fact, an optical isolation is required for the IM that is orders of magnitudes larger than the one for the PM. Even so, this difference is not larger than about 60~dB~\cite{tha3}. This roughly corresponds to the optical isolation displayed by an inexpensive commercially available component like a dual-stage optical isolator. Hence this solution is well within the feasibility range of current technology.

\subsection{Individual THA - Case~2}
\label{subsec:indTHA2}

In the previous section, we considered a worst-case assumption for the amount of light leaked out of the QKD transmitter, Eq.~(\ref{set1}). In that model, the leaked intensity was independent of the inner setting of the transmitter. On the one hand, this permits to bypass the precise characterisation of the QKD setup. On the other hand, it neglects a few physical considerations that can considerably improve the key rate. For example, the fraction of Eve's light that is back-reflected by a component that precedes the modulators in the transmitter's architecture does not contribute to the THA. A second important consideration is that, according to the initial worst-case scenario drawn for the individual THA, the modulators are driven with a NRZ logic. This entails that most of the time during the encoding process the modulators' medium is non-reflective, as its refractive index is homogeneous and constant between two consecutive NRZ modulation values. Hence, the THA has to be executed exploiting not the reflectivity of the IM (or PM), but that of the interfaces coming after it in the transmitter's architecture instead. Specifically, the THA would run as follows\footnote{We explicitly consider the IM in this description but the argument also applies to the PM.}: Eve's light passes through the IM a first time; it hits an interface placed after the IM and is reflected back from it towards the IM; it passes through the IM a second time and is finally leaked out of the QKD system into Eve's hands. During this two-way trip through the IM, Eve's light undergoes the same changes as the signals prepared by the transmitter for a normal QKD session. Therefore the leaked light is now highly informative of the inner settings of the transmitter.

In principle, a two-way round trip through a NRZ-modulated IM entails a double attenuation of Eve's light. However, because attenuation plays against Eve in a THA, it is conservative to assume that Eve's light is attenuated only once by the IM. To fix the ideas we can think that it passes unattenuated through the IM on the forward path and then is attenuated on the backward path in exactly the same way as the legitimate signals are. In this new scenario, the settings for the amplitudes of the leaked light are:
\begin{equation}\label{set2}
  \beta_{\rm \gamma_s}^2 = I_{\rm max}, \qquad \beta_{\rm \gamma_v}^2 = \frac{\gamma_{\rm v}}{\gamma_{\rm s}} I_{\rm max}, \qquad \beta_{\rm \gamma_w}^2 = \frac{\gamma_{\rm w}}{\gamma_{\rm s}} I_{\rm max}.
\end{equation}
Hence, differently from the previous case, Alice's modulation of the intensity directly affects now the information leaked to the eavesdropper for any fixed value of $I_{\rm max}$. The detailed calculation of the trace distance terms $D_{n,j,k,l}$ for the leaked states under Eq.~(\ref{set2}) is given in~\ref{subsec:appen_parD_Case2}.

\begin{figure}[h!]
\includegraphics[width=0.65\columnwidth]{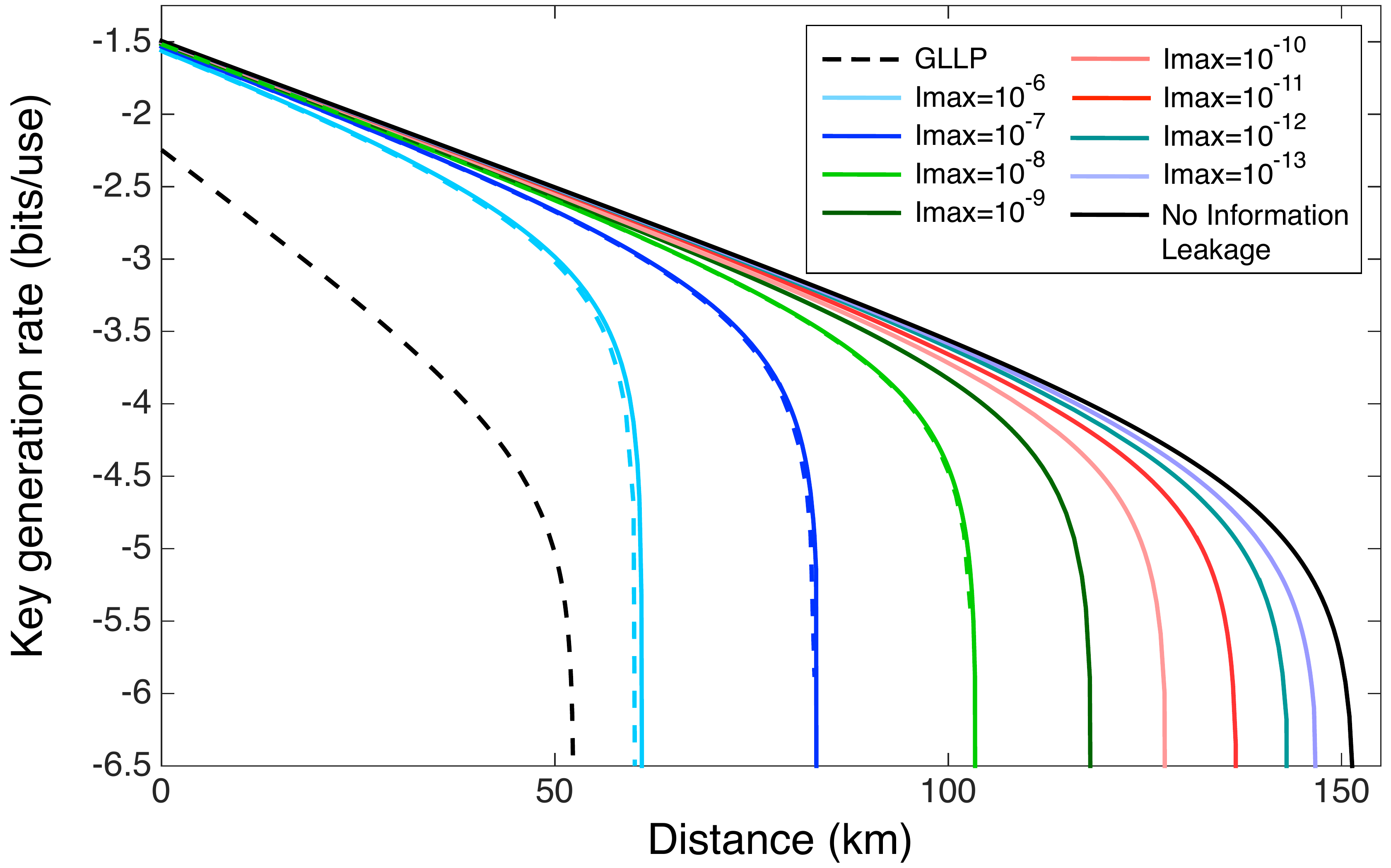}
\centering
\caption{Secure key rate versus distance in presence of a THA targeting the modulating devices of a QKD transmitter. Each colour corresponds to a different value of $I_{\rm max}$. The depicted key rate is obtained when the intensity of the leaked light is modulated in the same way as for the standard light pulses in decoy-state QKD, see Eq.~(\ref{set2}) in the main text. The solid lines are for a leakage due only to the IM, while the dashed lines, visible for $I_{\rm max}$ equal to $10^{-6}$, $10^{-7}$ and $10^{-8}$, correspond to the total leakage coming from the IM and the PM of the transmitter simultaneously. For every distance, the key rate is minimised over the angles $\theta_{\gamma_j}$, controlled by Eve, and maximised over the amplitudes ${\rm \gamma_s}$ and ${\rm \gamma_v}$, controlled by Alice. All the parameters used in the simulation are listed in Table~\ref{table1}.}
\label{fig:case2}
\end{figure}
In Fig.~\ref{fig:case2}, we plot the secure key rate as a function of the distance between the users, varying the parameter $I_{\rm max}$. The key rate has improved with respect to that in Fig.~\ref{fig:case1}. For the largest intensity of the leakage in the figure, $I_{\rm max}=10^{-6}$, the key rate derived from our security proof reaches about 60~km distance and is always better than the one attained with the GLLP approach. Moreover, for a leakage intensity $I_{\rm max}=10^{-12}$, the key rate is nearly indistinguishable from the rate of an ideally shielded system (black solid line in Fig.~\ref{fig:case2}) up to about 100~km, that is 70\% of the maximum transmission distance.

As in the previous case, we include in the figure the simulation of the key rate under a THA simultaneously run against the IM and the PM (dashed coloured lines in Fig.~\ref{fig:case2}). Again, the THA against the PM only marginally affects the overall key rate. Therefore the countermeasure to information leakage based on readily available optical isolators, discussed in the previous Sec.~\ref{subsec:indTHA1}, still applies here. Indeed, this solution becomes even more effective in the realistic scenario described in the present section, due to the better secure key rate shown in Fig.~\ref{fig:case2} in comparison with the worst-case key rate presented in Fig.~\ref{fig:case1}.

\subsection{Individual THA - Case~3}
\label{subsec:indTHA3}

In this section, we further improve the key rate under a THA by considering the phase randomisation of Eve's signal, as discussed in Sec.~\ref{Sec: Trace distance}. Phase randomisation can drastically reduce the dangerousness of the THA as it removes any residual entanglement with the eavesdropper's probes, and we can expect higher key with the phase randomisation due to the non-existence of the off-diagonal elements.   However, on the other hand, one has to be very careful about how phase randomisation is implemented, as this could open new loopholes. For example, if it is realised by adding a supplementary modulator to the system, Eve could first direct the THA against this new device to learn the phase information, and then address the PM and the IM as in the non-phase-randomised case, thus suppressing all the benefits due to the randomisation of the phase.

However, we showed in the previous sections that the THA against the PM is less effective than the one against the IM. Therefore, in order to improve the performance of the system against the THA, it is more important to randomise the phase of Eve's light directed against the IM than the one against the PM. This offers an alternative, possibly more robust, way to implement phase randomisation. Specifically, we can avoid using an additional ad-hoc module and focus rather on the working mechanism of the IM, which is part already of the transmitting unit. A common technique to modulate intensity is via a symmetric Mach-Zehnder interferometer (MZI). The light entering the MZI is first split into two beams and then recombined with a suitable phase. This will generate interference and therefore intensity modulation at the output ports of the MZI. By blocking one of the output ports, intensity modulation is obtained from the unblocked port as a result of the destructive or constructive interfering process. To modulate the relative phase between the two arms of the MZI, it is sufficient to control the refractive index of only one of the two MZI arms, and this is the most commonly used configuration. However, if a `dual-drive' IM is used instead, both the arms in the MZI can be independently controlled, so to gain simultaneous control over the relative phase as well as the global phase of the signals traversing the IM respect to an external reference phase. If the global phase in the dual-drive IM is randomised, by encoding in each time slot a different phase value, Eve's probing signal will be phase randomised too and its phase will become uninformative to Eve. In this case, the state of the leaked signals seen by Eve will not be the one in Eq.~(\ref{coh-state}) any more and will be replaced by the following one:
\begin{equation}\label{set3}
  \rho_{\gamma_j} = e^{-\beta_{\gamma_j}^2} \sum_{n=0}^{\infty}\frac{\beta_{\gamma_j}^{2n}}{n!}|n\rangle\langle n|.
\end{equation}
We simulate the secure key rate for this situation using the detailed calculation of the trace distance terms $D_{n,j,k,l}$ given in~\ref{subsec:appen_parD_Case3} and setting the intensities $\beta_{\gamma_j}^2$ as in Eq.~(\ref{set2}). That is, we still consider a THA where Eve's light crosses the IM first and is back-reflected to Eve from an interface placed after the IM in the transmitter architecture.

\begin{figure}[h!]
\includegraphics[width=0.65\columnwidth]{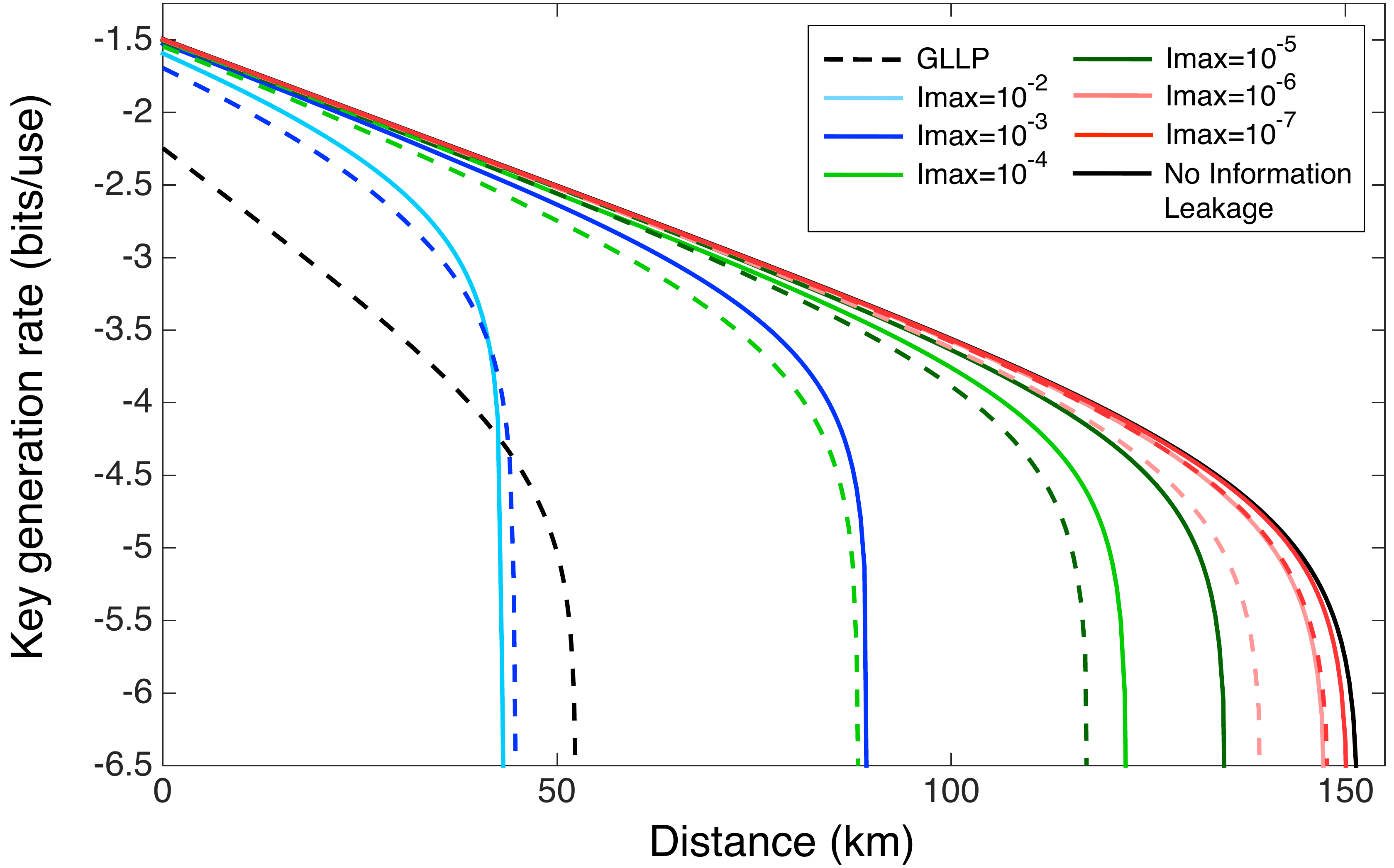}
\centering
\caption{Secure key rate versus distance in presence of a THA targeting the modulating devices of a QKD transmitter. Each colour corresponds to a different value of the leaked intensity $I_{\rm max}$. The phase of the leaked light is randomised, see Eq.~(\ref{set3}) in the main text. The solid lines are for a leakage due only to the IM while the dashed lines, corresponding to a much lower rate, are for the total leakage due to IM and PM simultaneously. For every distance, the key rate is maximised over the amplitudes ${\rm \gamma_s}$ and ${\rm \gamma_v}$. All the parameters used in the simulation are listed in Table~\ref{table1}. }
\label{fig:case3}
\end{figure}
The result of the simulation is reported in Fig.~\ref{fig:case3}. It is apparent that the key rate has vastly improved with respect to Figs.~\ref{fig:case1} and \ref{fig:case2}. Even for a leakage intensity as large as $I_{\rm max}=10^{-2}$, the key rate remains positive up to about 40~km. For an intensity  smaller than $I_{\rm max}=10^{-6}$, the resulting key rate is indistinguishable from the ideal one (solid black line) over almost the whole distance range. This shows the beneficial effect of phase randomisation, which was expected from the discussion in Sec. 3.1.1. Differently from previous cases, the simultaneous information leakage from IM and PM (dashed lines in the figure) leads now to a key rate that is apparently lower than for a leakage due to the IM only (solid lines in the figure). For example, when $I_{\rm max}=10^{-3}$ and the leakage is due to the IM only, the key rate remains positive for more than 85~km, while it falls below 50~km for a simultaneous leakage from PM and IM.

Given the benefit of phase randomisation, a natural question arises if sending
only an $n$-photon Fock state, rather than its classical mixture, is beneficial to Eve. In~\ref{lastq}, we discuss this point, and we show that the benefit of employing this attack Eve
obtains is negligibly small, {\it i.e.,} this attack can enlarge the trace distance only by the order of the transmission rate of Alice's device, which is negligible given the proper installation of optical isolators and filters. By recalling that phase randomisation transforms {\it any} state into a classical mixture of Fock states, we can conclude that the results presented in this section are essentially the secure key rate with the most general THA against IM assuming the phase randomisation.

From a practical perspective, phase randomisation makes the IM as robust against information leakage as a non-phase-randomised PM. This, in combination with the enhanced security due to the removal of any residual entanglement with Eve as well as that of all the off-diagonal elements, promotes phase randomisation as a relevant countermeasure to prevent the THA and the information leakage in general from the transmitter of decoy-state QKD and mdiQKD.

\begin{figure}[h!]
\includegraphics[width=0.85\columnwidth]{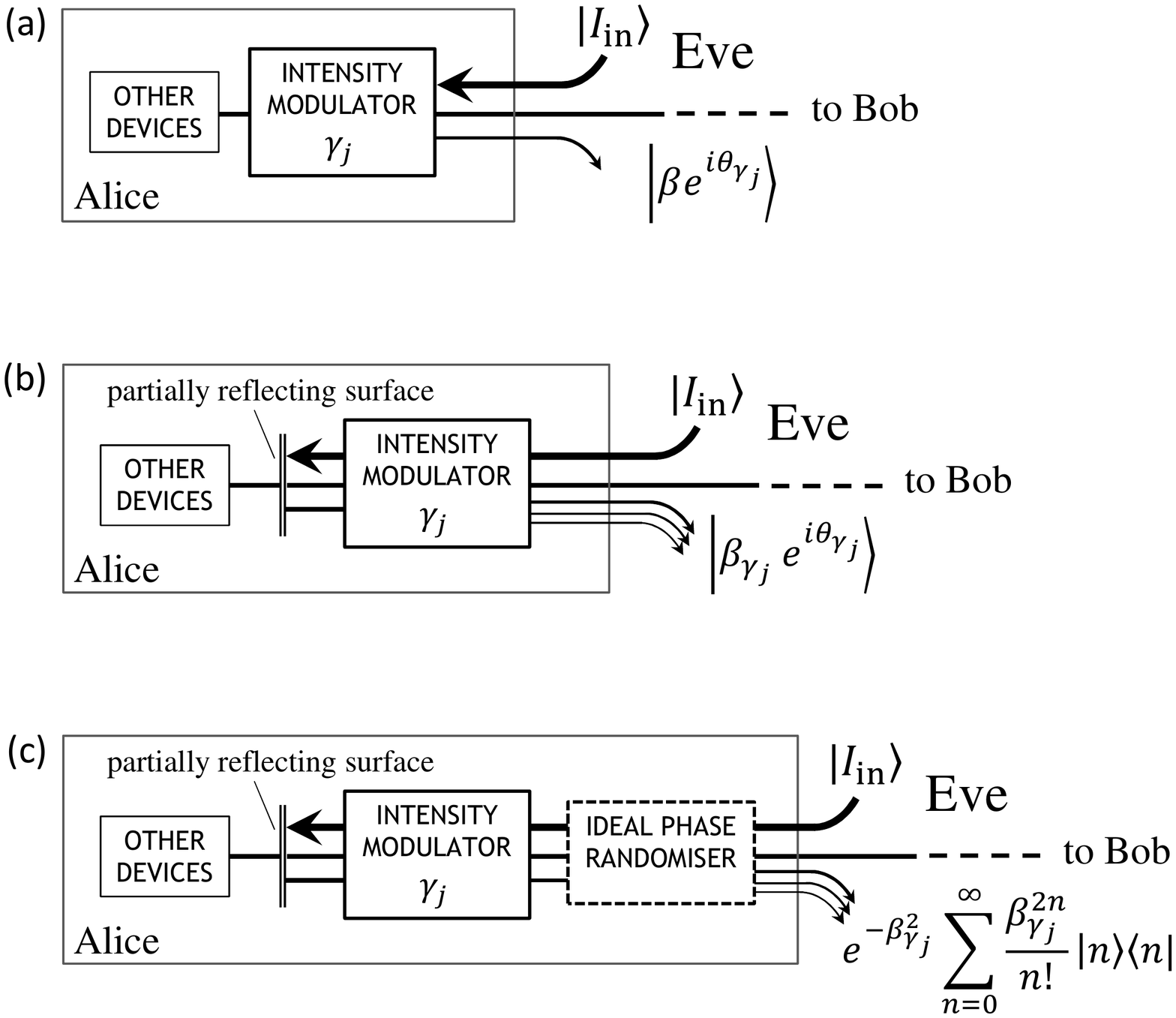}
\centering
\caption{Models and states used in the simulation of individual THA targeting the IM. (a) The intensity of the back-reflected light is independent of the intensity settings (Case 1, Sec.~\ref{subsec:indTHA1}). (b) The intensity of the back-reflected light depends on the intensity settings (Case 2, Sec.~\ref{subsec:indTHA2}). (c) Eve's back-reflected light is phase randomised and it is represented by the classical mixture of Fock states (Case 3, Sec.~\ref{subsec:indTHA3}).}
\label{fig:models}
\end{figure}

Before concluding this section, it is useful to summarise the physical models and the assumptions underlying our simulations. This is done with the help of Fig.~\ref{fig:models}.

In Sec.~\ref{subsec:indTHA1}, we considered the scenario depicted in Fig.~\ref{fig:models}(a), leading to Eq.~(\ref{set1}). In this case, the coherent state of light back-reflected by the IM carries in its phase $\theta_{\gamma_j}$ the information about the intensity settings of the IM, $\gamma_j$. Its amplitude is the maximum allowed by any physical mechanism used to limit Eve's input light, irrespective of the IM settings, thus making the states outputted by Alice more distinguishable to Eve. This, together with the choice of the angles $\theta_{\gamma_j}$, chosen to be most favourable to Eve, let us draw the worst-case key rate lines shown in Fig.~\ref{fig:case1}.

In Sec.~\ref{subsec:indTHA2}, we devised a more realistic scenario, depicted in Fig.~\ref{fig:models}(b). Typically, Eve's light is reflected by an interface placed after the IM (double line in the figure) rather than by the IM itself. During the THA, Eve's light can pass through the IM when it is fully transmissive, to be reflected by the interface and pass through the IM again when the intensity settings are on. This way, Eve's light is modulated with exactly the same settings $\gamma_j$ used by Alice for her own states, leading to Eq.~(\ref{set2}) and Fig.~\ref{fig:case2}.

Finally, in Sec.~\ref{subsec:indTHA3}, we applied phase randomisation to any light emerging from Alice's module, as shown in Fig.~\ref{fig:models}(c). For the intensity of the light back-reflected to Eve, we considered the same scenario as in Fig.~\ref{fig:models}(b). The ideal phase randomiser shown in the figure is a powerful resource as it removes any phase information from the output states, see Eq.~(\ref{set3}), leading to better key rates, as reported in Fig.~\ref{fig:case3}.

The model used to draw the lines for the PM is not explicitly described, as it is similar to the IM one and is detailed in \cite{tha3}. Although the cases described in this section do not constitute an exhaustive list, they represent useful practical cases and can be used as guidelines for the secure implementation of real QKD systems.

\section{Discussion}\label{discussion}

In this work, we have presented a general formalism to calculate the secret key rate of decoy-state QKD and mdiQKD under \textit{any} THA directed against the transmitter's modulators.
It is useful to give some insight into this formalism, in particular the one for IM, and discuss why the THA affects the standard theory of decoy states.

In the analysis of the decoy-state method without the THA, a fictitious protocol is considered where Alice delays her decision on the intensity settings after Bob detects a pulse. That is, after the detection of the pulse, Alice randomly decides the intensity setting $\gamma_{\rm s}$, $\gamma_{\rm v}$ and $\gamma_{\rm w}$~\cite{mdiQKDfinite}.
For simplicity, let us consider the discrimination between the signal state ${\rm s}$ and the first decoy state ${\rm v}$ only. By using the relation ${\rm Pr}(\gamma_{\rm s}^{i}|{\rm click}, n)+{\rm Pr}(\gamma_{\rm v}^{i}|{\rm click}, n)=1$ and the Bayes's rule, we can rewrite Eq.~(\ref{basis-eq-THA}) as:
\begin{eqnarray}
\left|\frac{{\rm Pr}(\gamma_{\rm v})}{{\rm Pr}(\gamma_{\rm s})+{\rm Pr}(\gamma_{\rm v})}- {\rm Pr}(\gamma_{\rm v}^{i}|{\rm click}, n)\right|\le \frac{{\rm Pr}(\gamma_{\rm s}){\rm Pr}(\gamma_{\rm v})}{{\rm Pr}(\gamma_{\rm s})+{\rm Pr}(\gamma_{\rm v})} \frac{{\overline D_{n, {\rm s}, {\rm v}}}}{{\rm Pr}^{i}({\rm click}|n)}\,,
\label{delay-THA2}
\end{eqnarray}
where ${\overline D_{n, {\rm s}, {\rm v}}}:={\rm max}_{i} D^{i}_{n, {\rm s}, {\rm v}}$. From this equation, we see that in the case of no THA, ${\overline D_{n, {\rm s}, {\rm v}}}=0$ and $\frac{{\rm Pr}(\gamma_{\rm v})}{{\rm Pr}(\gamma_{\rm s})+{\rm Pr}(\gamma_{\rm v})}= {\rm Pr}(\gamma_{\rm v}^{i}|{\rm click}, n)$ for any $i$ and $n$. This entails that Alice's assignment of the intensities in the fictitious protocol can be made identical and independent of the instance $i$ over the detected instances. This allows us to use probability inequalities, such as the Multiplicative Chernoff bound~\cite{mdiQKDfinite}, which applies to independent trials. However, when the THA is on the line, ${\overline D_{n, {\rm s}, {\rm v}}}\neq0$ and the bound to the L.H.S. of Eq.~(\ref{delay-THA2}) becomes dependent on the instance $i$. To solve this problem, we make use of Azuma's inequality~\cite{azuma}. Because we are in the asymptotic scenario, the technical details related to the inequality are unnecessary and we will not write them here explicitly.

Our formalism does not require any knowledge of Eve's measurement for the THA or the detailed specification of the state used. Instead, a detailed characterisation of the modulators is needed. This is important because while the full characterisation of Alice's modulators over many modes is doable at least in principle, the characterisation of Eve's THA is impossible even in principle. However, the full characterisation of Alice's modulators might be challenging in practice and further
research needs to be done in this direction. We remark that our formalism is a powerful tool in this context because it can readily accept any
mathematical model that describes the behaviour of the modulators.

As we have discussed in Sec.~\ref{Sec: Trace distance} and practically demonstrated in Sec.~\ref{subsec:indTHA3}, it is important to perform phase randomisation of Eve's signals to defeat the THA exploiting entanglement and to enhance the key rate. However, on the other hand, it is important to perform the randomisation without opening additional loopholes. Also, the question remains of whether this solution is more practical than the one based on a series of optical isolators. Active phase randomisation requires precise synchronisation and a sequence of random numbers in the input. Even if correctly performed, a certain level of optical isolation is always needed to shield a system from the external environment. The total amount of required isolation clearly depends on phase randomisation, as seen by comparing Figs.~\ref{fig:case2} and \ref{fig:case3}. However, these figures also show that the difference in the values of $I_{\rm max}$ amounts roughly to 60~dB, which can be achieved with a single entirely passive component like a dual-stage optical isolator. Hence even high isolation levels can be inexpensively achieved through a series of such isolators.

\section{Conclusion}\label{concl}

In this paper, we have quantified the secure key rate of decoy-state-based QKD in presence of leaky transmitters. This allowed us to suggest quantitative countermeasures to restore security even in this more general scenario. A real setup is typically leaky in practice, due to the presence of side channels hidden in the preparation of the communication signals, or due to the active intervention of an eavesdropper. The analysis of this case is then of immediate practical interest.
Our analysis applies to any decoy-state system that uses an intensity modulator or a phase modulator to distill a quantum key. It includes in fact the most general attack based on the extra information possibly leaked from such devices.

We have employed our formalism to analyse particular examples of THA, where Eve exploits coherent states of light to probe the intensity and phase modulators in the transmitter. Our results show that it is possible to distill a key from leaky transmitters that approach the ideal rate of a perfectly shielded system. For that, two main solutions play a crucial role. On one hand, optical isolation has to be guaranteed for any system through an adequate number of attenuators and optical isolators. On the other hand, active phase randomisation can further enhance the protection, removing any residual entanglement from Eve's probing signals.

Given the generality of our approach and its applicability to cases of practical interest, we believe that it will become a fundamental tool to analyse the security of real-world quantum communication systems, including those for standard QKD, mdiQKD and the device-independent QKD where phase modulators and/or intensity modulators are used.

\section*{Acknowledgements}

The authors wish to thank Norbert L\"{u}tkenhaus for very useful discussions, and the anonymous referees for their constructive comments which have helped us to
significantly improve the contents of this paper.
This work was supported by the Galician Regional Government (program ``Ayudas para proyectos de investigaci\'on desarrollados por investigadores emergentes'' EM2014/033, and consolidation of Research Units: AtlantTIC), the Spanish Ministry of Economy and Competitiveness (MINECO), the ÒFondo Europeo de Desarrollo RegionalÓ (FEDER) through grant TEC2014-54898-R, the ImPACT Program of the Council for Science, Technology and Innovation (Cabinet Office, Government of Japan), and the project EMPIR 14IND05 MIQC2. This project has received funding from the EMPIR programme co-financed by the Participating States and from the European Union's Horizon 2020 research and innovation programme.

\appendix

\section{Azuma's inequality}\label{Explanation of Azuma}
In this Appendix we introduce Azuma's inequality~\cite{azuma}. It can be applied to a sequence of random variables
$X^{(0)}, X^{(1)},...,X^{(l)}$ that satisfies the Martingale and the Bounded difference conditions (BDC).
In particular, a set of random variables is called a Martingale if and only if
$E[X^{(l+1)}|X^{(0)}, X^{(1)},...,X^{(l)}]=X^{(l)}$ holds for any $l$, where $E[\cdot]$ represents the expectation value. That is, the expectation value of the $(l+1)^{th}$ random variable conditional on all
the previous random variables is equal to the $l^{th}$ random variable. On the other hand, $X^{(0)}, X^{(1)},...,X^{(l)}$ satisfies the BDC if and only if there exists $c^{(l)}>0$ such that
$|X^{(l+1)}-X^{(l)}|\leq{}c^{(l)}$ for any $l$. In this scenario, Azuma's inequality states that
\begin{equation}
{\rm Pr}[\left|X^{(l)}-X^{(0)}\right|>l\delta]\leq2e^{-\frac{l^2\delta^2}{2\sum_{k=1}^{l}(c^{(l)})^2}}
\end{equation}
for any $\delta\in(0,1)$.

Now, to derive the result that we use in the main text, we proceed as follows. In particular, let us consider that we flip coins starting from the first coin in order. The coins can be correlated in an arbitrary manner. Let $y_{u}$ be the random variable that represents the result of the $u^{th}$ coin, with $y_{u}=1$ when the result is head and $y_{u}=0$ when it is tail. Let $P(y_{u}=1|\xi_0,...,\xi_{u-1})$ be the conditional probability of having head in the $u^{th}$ coin
conditional on all the results of the previous coins, which we denote as $\xi_0,...,\xi_{u-1}$.
Finally, we denote by  $\Lambda^{(l)}$ the actual number of heads obtained after flipping $l$ coins. Then, it can be shown
that
\begin{equation}
X^{(l)}:=\Lambda^{(l)}-\sum_{u=1}^lP(y_{u}=1|\xi_0,...,\xi_{u-1}).
\end{equation}
is a Martingale and satisfies the BDC. We have, therefore, that
\begin{equation}
{\rm Pr}[|\Lambda^{(N)}-\sum_{u=1}^NP(y_{u}=1|\xi_0,...,\xi_{u-1})|>N\delta]\leq{2e^{-\frac{N\delta^2}{2}}}.
\label{Azumainequality}
\end{equation}

\section{Joint THA when the IM and the PM are correlated}\label{Correlation}
In this Appendix, we explain briefly how to adapt our formalism to evaluate as well the situation where there are arbitrary
correlations between the IM and the PM, and Eve can exploit this fact in her THA.

In this correlated scenario, Alice and Bob could first estimate
the bit and basis dependent single-photon yield, which we denote as
$Y_{1}^{\gamma_{\rm s}, \xi_{\rm A}, \zeta_{\rm A}}$, for the signal setting. Here, the parameter $\xi_{\rm A}$ denotes Alice's bit value and $\zeta_{\rm A}$
is her basis choice. That is, $Y_{1}^{\gamma_{\rm s}, \xi_{\rm A}, \zeta_{\rm A}}$ represents the conditional probability that Bob obtains a ``click'' event
given that Alice selects the signal setting and sends him a single-photon state encoding a bit value $\xi_{\rm A}$ in the basis $\zeta_{\rm A}$.
To estimate this yield, Alice can declare Bob (over the authenticated public channel) all the bit and basis information associated
to those instances where she used a decoy setting and Bob obtained a ``click'' event.
With this information, Alice and Bob can estimate $Y_{1}^{\gamma_{\rm s}, \xi_{\rm A}, \zeta_{\rm A}}$
by using a modified version of Eq.~(\ref{basis-eq-THA3}) given by
\begin{eqnarray}
\left|Y_n^{\gamma_j,\xi_{\rm A}, \zeta_{\rm A}}-\left[q_{nkl}Y_n^{\gamma_k,\xi_{\rm A}, \zeta_{\rm A}}+(1-q_{nkl})Y_n^{\gamma_l,\xi_{\rm A}, \zeta_{\rm A}}\right]\right|&\le& D^{\xi_{\rm A}, \zeta_{\rm A}}_{n, j, k,l}\,.
\end{eqnarray}
Here, the parameter $D^{\xi_{\rm A}, \zeta_{\rm A}}_{n, j, k,l}$ is a modified version of $D_{n, j, k,l}$ that refers
solely to the set of choices $\{\xi_{\rm A}, \zeta_{\rm A}\}$. Note that this modification is needed because now the
decoy-state method is bit-and-basis-dependent.
Similarly, one can obtain the single-photon bit error rate for the signal setting
by estimating the yields $Y_{1}^{\gamma_{\rm s}, \xi_{\rm A}, \zeta_{\rm A}, \xi_{\rm B}, \zeta_{\rm B}}$.
Here, $\xi_{\rm B}$ denotes the bit value obtained by Bob and $\zeta_{\rm B}$
is his measurement basis choice. That is, 
$Y_{1}^{\gamma_{\rm s}, \xi_{\rm A}, \zeta_{\rm A}, \xi_{\rm B}, \zeta_{\rm B}}$
is the conditional probability that Bob obtains the bit value $\xi_{\rm B}$ conditioned on the fact that Alice emits a single-photon pulse that encodes the bit value $\xi_{\rm A}$ in the $\zeta_{\rm A}$ basis with the setting $\gamma_{\rm s}$ and Bob chooses the $\zeta_{\rm B}$ basis for his measurement.
In other words, note that $Y_{1}^{\gamma_{\rm s}, \xi_{\rm A}, \zeta_{\rm A}, \xi_{\rm B}, \zeta_{\rm B}}$ also depends on Bob's basis choice and on his bit value.

After obtaining the single-photon yield as well as the associated error rate,
Alice and Bob generate a secret key from those instances
where Alice emitted a single-photon pulse prepared in the Z basis and using the signal setting, and Bob obtained  a ``click'' event when he measured the pulse in the Z basis.
Note that all the statistics associated to such instances
are estimated through the bit-and-basis-dependent decoy-state method.
Likewise, one can readily obtain the single-photon yield
associated to those events where Alice emits a
single-photon pulse prepared in the X basis and using the signal setting, and Bob obtains a ``click'' event when he uses the X basis.
Now, to generate
a key, one follows the technique explained in Sec.~\ref{subsec:THA against the PM} except from a small modification. In particular, one has to
replace ${\hat U}_{\rm A_{p}, A_{a}, E_{p}}^{\zeta, i}$ in Eq.~(\ref{Unitary for THA PM}) with ${\hat U}_{\rm A_{p}, A_{a}, E_{p}}^{\zeta, i, \gamma_{\rm s}, 1}$,
which is a restricted version of ${\hat U}_{\rm A_{p}, A_{a}, E_{p}}^{\zeta, i}$ that only considers
the single-photon emission part in the signal setting. That is, Alice and Bob have to
characterise the behaviour of the PM depending on their bases choice when they select the signal setting.

With the modifications above, one can obtain the phase error rate $\delta_{{\rm X}-{\rm Error}|{\rm Z}}$ from Eq.~(\ref{auto})
because the bit-and-basis-dependent decoy-state method allows us to evaluate all the parameters needed to solve this equation, all of which are now restricted only to
the single-photon emission events.
Then, in the asymptotic limit of a large number of transmitted signals, we have that the secure key rate 
is given by
\begin{eqnarray}\label{casa}
  K &\geq& \sum_{i=0}^1q \{p_0^{\gamma_{\rm s}, i, {\rm Z}} Y_{0{\rm L}}^{\gamma_{\rm s}, i, {\rm Z}}
  + p_1^{\gamma_{\rm s}, i, {\rm Z}} Y_{1{\rm L}}^{\gamma_{\rm s}, i, {\rm Z}}
[ 1-h(\delta_{{\rm X}-{\rm Error}|{\rm Z}}) ] \nonumber\\
&-&f(E^{\gamma_{\rm s}}) Q^{\gamma_{\rm s}} h(E^{\gamma_{\rm s}})\},
\end{eqnarray}
where $p_0^{\gamma_{\rm s}, i, {\rm Z}}$ is the probability that Alice emits
the vacuum state given that she chooses $\gamma_{\rm s}$ and the bit value $i$ in the Z basis. The other
parameters which appear in Eq.~(\ref{casa}) are defined in a similar manner (see also Eq.~(\ref{rate-eq})). 
Therefore, we conclude that one can apply our formalism to analyse also the case where
there are arbitrary correlations between the IM and the PM, and prove security in the most general case, given that 
a full description of the behaviour of these two devices is available.

\section{Estimation of $Y_{0{\rm L}}^{\gamma_{\rm s}}$, $Y_{1{\rm L}}^{\gamma_{\rm s}}$ and $e_{1{\rm U}}^{\gamma_{\rm s}}$}
\label{app_est}

In this Appendix we show that these parameters can be estimated using linear programming.
Such instances of optimisation problems can be solved efficiently in polynomial time~\cite{lp}.
Although the estimation method presented here is valid for any number of decoy states used by Alice, we will assume, like in the main text, that Alice employs three different intensity settings: $\gamma_{\rm s}$, $\gamma_{\rm v}$ and $\gamma_{\rm w}$.

Our starting point is Eq.~(\ref{basis-eq-THA3}). Let us consider first the case $k=l$.
As shown in~\ref{appen_parD}, the parameters $D_{n, j, k}$ do not depend on the photon number $n$, at least for the examples considered in Sec.~\ref{simulations}. This means, in particular, that this equation can be rewritten as
\begin{eqnarray}\label{basis-eq-THA3c}
\left|Y_n^{\gamma_j}-Y_n^{\gamma_k}\right|&\le& D_{j, k},
\end{eqnarray}
or, equivalently, the yields $Y_n^{\gamma_j}$ and $Y_n^{\gamma_k}$ satisfy
\begin{equation}
Y_n^{\gamma_j}=Y_n^{\gamma_k}+\Delta^{jk},
\end{equation}
with $\Delta^{jk}\in[-D_{j, k},D_{j, k}]$. Since Alice uses three different intensity settings, we have the following six conditions
\begin{eqnarray}
&& Y_n^{\gamma_{\rm s}}=Y_n^{\gamma_{\rm v}}+\Delta^{{\rm s}{\rm v}}, \quad \quad \quad Y_n^{\gamma_{\rm s}}=Y_n^{\gamma_{\rm w}}+\Delta^{{\rm s}{\rm w}}, \nonumber \\
&& Y_n^{\gamma_{\rm v}}=Y_n^{\gamma_{\rm s}}+\Delta^{{\rm v}{\rm s}}, \quad \quad \quad Y_n^{\gamma_{\rm v}}=Y_n^{\gamma_{\rm w}}+\Delta^{{\rm v}{\rm w}}, \nonumber \\
&& Y_n^{\gamma_{\rm w}}=Y_n^{\gamma_{\rm s}}+\Delta^{{\rm w}{\rm s}}, \quad \quad \quad Y_n^{\gamma_{\rm w}}=Y_n^{\gamma_{\rm v}}+\Delta^{{\rm w}{\rm v}}.
\end{eqnarray}
By combining the first and the third one, we find, for example, that $\Delta^{{\rm s}{\rm v}}=-\Delta^{{\rm v}{\rm s}}$. Similarly, we obtain $\Delta^{{\rm s}{\rm w}}=-\Delta^{{\rm w}{\rm s}}$ and $\Delta^{{\rm v}{\rm s}}-\Delta^{{\rm w}{\rm s}}=\Delta^{{\rm v}{\rm w}}=-\Delta^{{\rm w}{\rm v}}$. By using this last condition we find, therefore, that
\begin{eqnarray}\label{eqq1}
&& Y_n^{\gamma_{\rm v}}=Y_n^{\gamma_{\rm s}}+\Delta^{{\rm v}{\rm s}}=Y_n^{\gamma_{\rm s}}+\Delta^{{\rm w}{\rm s}}+\Delta^{{\rm v}{\rm w}}, \nonumber \\
&& Y_n^{\gamma_{\rm w}}=Y_n^{\gamma_{\rm s}}+\Delta^{{\rm w}{\rm s}}.
\end{eqnarray}
That is, we can express the yields $Y_n^{\gamma_{\rm v}}$ and $Y_n^{\gamma_{\rm w}}$ as a function of $Y_n^{\gamma_{\rm s}}$ and the parameters $\Delta^{{\rm w}{\rm s}}$ and $\Delta^{{\rm v}{\rm w}}$.

Next, we consider the case $k \neq l$. In this scenario, Eq.~(\ref{basis-eq-THA3}) can be rewritten as
\begin{eqnarray}\label{basis-eq-THA3b}
Y_n^{\gamma_j}=q_{nkl}Y_n^{\gamma_k}+(1-q_{nkl})Y_n^{\gamma_l}+\Delta^{njkl},
\end{eqnarray}
for all $n$, where $\Delta^{njkl}\in[-D_{n,j, k, l},D_{n, j, k, l}]$. We have, therefore, the following three conditions:
\begin{eqnarray}\label{basis-eq-THA3d}
Y_n^{\gamma_{\rm s}}&=&q_{n{\rm vw}}Y_n^{\gamma_{\rm v}}+(1-q_{n{\rm vw}})
Y_n^{\gamma_{\rm w}}+\Delta^{n{\rm svw}}, \nonumber \\
Y_n^{\gamma_{\rm v}}&=&q_{n{\rm sw}}Y_n^{\gamma_{\rm s}}+(1-q_{n{\rm sw}})
Y_n^{\gamma_{\rm w}}+\Delta^{n{\rm vsw}}, \nonumber \\
Y_n^{\gamma_{\rm w}}&=&q_{n{\rm sv}}Y_n^{\gamma_{\rm s}}+
(1-q_{n{\rm sv}})
Y_n^{\gamma_{\rm v}}+\Delta^{n{\rm wsv}}.
\end{eqnarray}
If we substitute in these equations the value of $Y_n^{\gamma_{\rm v}}$ and $Y_n^{\gamma_{\rm w}}$ given by Eq.~(\ref{eqq1}) we obtain
the following three equality constraints:
\begin{eqnarray}\label{basis-eq-THA3e}
0&=&\Delta^{{\rm w}{\rm s}}+q_{n{\rm vw}}\Delta^{{\rm v}{\rm w}}+\Delta^{n{\rm svw}}, \nonumber \\
0&=&q_{n{\rm sw}}\Delta^{{\rm w}{\rm s}}+\Delta^{{\rm v}{\rm w}}-\Delta^{n{\rm vsw}}, \nonumber \\
0&=&q_{n{\rm sv}}\Delta^{{\rm w}{\rm s}}-
(1-q_{n{\rm sv}})\Delta^{{\rm v}{\rm w}}-\Delta^{n{\rm wsv}}.
\end{eqnarray}
Finally, by taking into account that $\Delta^{njkl}\in[-D_{n,j, k, l},D_{n, j, k, l}]$ for all $n$ and for all $j,k,l\in\{{\rm s, v, w}\}$ with $j\neq k\neq l$, we have that
to satisfy Eq.~(\ref{basis-eq-THA3e}) we must fulfill the following conditions:
\begin{eqnarray}\label{basis-eq-THA3f}
-D_{n,{\rm s,v,w}}&\le&\Delta^{{\rm w}{\rm s}}+q_{n{\rm vw}}\Delta^{{\rm v}{\rm w}}\le D_{n,{\rm s,v,w}}, \nonumber \\
-D_{n,{\rm v,s,w}}&\le&q_{n{\rm sw}}\Delta^{{\rm w}{\rm s}}+\Delta^{{\rm v}{\rm w}}\le D_{n,{\rm v,s,w}}, \nonumber \\
-D_{n,{\rm w,s,v}}&\le&
q_{n{\rm sv}}\Delta^{{\rm w}{\rm s}}-(1-q_{n{\rm sv}})\Delta^{{\rm v}{\rm w}}
\le D_{n,{\rm w,s,v}}.
\end{eqnarray}

\subsection{Estimation of $Y_{0{\rm L}}^{\gamma_{\rm s}}$}

Here we present a linear program to estimate the parameter $Y_{0{\rm L}}^{\gamma_{\rm s}}$. We will assume that all the quantities below refer to events where both Alice and Bob use the same basis ({\it e.g.}, the Z basis), which will be considered as the key generation basis. We start by calculating the gain associated to the different intensity settings selected by Alice in this scenario. If we combine Eqs.~(\ref{gain_error}) and (\ref{eqq1}) we have that
\begin{eqnarray}
\nonumber  Q^{\gamma_{\rm s}} &=& \sum_{n=0}^\infty p^{\gamma_{\rm s}}_{n}Y_n^{\gamma_{\rm s}}, \\
Q^{\gamma_{\rm v}} &=& \sum_{n=0}^\infty p^{\gamma_{\rm v}}_{n}\left(Y_n^{\gamma_{\rm s}}+\Delta^{{\rm w}{\rm s}}+\Delta^{{\rm v}{\rm w}}\right)=\sum_{n=0}^\infty p^{\gamma_{\rm v}}_{n}Y_n^{\gamma_{\rm s}}+\Delta^{{\rm w}{\rm s}}+\Delta^{{\rm v}{\rm w}}, \nonumber \\
Q^{\gamma_{\rm w}} &=& \sum_{n=0}^\infty p^{\gamma_{\rm w}}_{n}\left(Y_n^{\gamma_{\rm s}}+\Delta^{{\rm w}{\rm s}}\right)=\sum_{n=0}^\infty p^{\gamma_{\rm w}}_{n}Y_n^{\gamma_{\rm s}}+\Delta^{{\rm w}{\rm s}}.
\label{gain_errorb}
\end{eqnarray}
That is, all the gains can be written as a function of the yields $Y_n^{\gamma_{\rm s}}$ together with the additional terms $\Delta^{{\rm w}{\rm s}}$ and $\Delta^{{\rm v}{\rm w}}$.

Eq.~(\ref{gain_errorb}) contains an infinite number of unknown parameters $Y_n^{\gamma_{\rm s}}$. Next, we reduce it to a finite set. For this, we derive a lower and upper bound for the gains $Q^\gamma$ that only depend on a finite number, ${\rm S_{cut}}+1$, of yields $Y_n^{\gamma_{\rm s}}$.
In particular, since $0\leq Y_n^{\gamma_{\rm s}}\leq 1$ and $p^{\gamma_{\rm s}}_{n}\geq 0$ for all $n$,
we have that
\begin{eqnarray}\label{sat_eve}
\nonumber  Q^{\gamma_{\rm s}} &\geq& \sum_{n=0}^{\rm S_{cut}} p^{\gamma_{\rm s}}_{n}Y_n^{\gamma_{\rm s}}, \\
Q^{\gamma_{\rm s}}&\leq& \sum_{n=0}^{\rm S_{cut}} p^{\gamma_{\rm s}}_{n}Y_n^{\gamma_{\rm s}}+\sum_{n={\rm S_{cut}}+1}^\infty p^{\gamma_{\rm s}}_{n}
=\sum_{n=0}^{\rm S_{cut}} p^{\gamma_{\rm s}}_{n}Y_n^{\gamma_{\rm s}}+\Gamma^{\gamma_{\rm s}},
\end{eqnarray}
for any ${\rm S_{cut}}\geq 0$. Here the parameter $\Gamma^{\gamma_{\rm s}}$ is defined as $\Gamma^{\gamma_{\rm s}}=\sum_{n={\rm S_{cut}}+1}^\infty p^{\gamma_{\rm s}}_{n}=1-\sum_{n=0}^{\rm S_{cut}} p^{\gamma_{\rm s}}_{n}$.
By using a similar procedure, one can obtain as well a lower and upper bound for $Q^{\gamma_{\rm v}}$ and $Q^{\gamma_{\rm w}}$.

Based on the foregoing, we find that $Y_{0{\rm L}}^{\gamma_{\rm s}}$ can be calculated using the following linear program:
\begin{eqnarray}\label{eq_solqqb}
\min \quad&&  Y_0^{\gamma_{\rm s}} \nonumber \\
{\rm s. t.}\quad&& Q^{\gamma_{\rm s}}\geq\sum_{n=0}^{\rm S_{cut}} p_n^{\gamma_{\rm s}} Y_n^{\gamma_{\rm s}}, \nonumber \\
&& Q^{\gamma_{\rm s}}-\Gamma^{\gamma_{\rm s}}\leq\sum_{n=0}^{\rm S_{cut}} p_n^{\gamma_{\rm s}} Y_n^{\gamma_{\rm s}}, \nonumber \\
&& Q^{\gamma_{\rm v}}\geq\sum_{n=0}^{\rm S_{cut}} p_n^{\gamma_{\rm v}} Y_n^{\gamma_{\rm s}}+\Delta^{{\rm w}{\rm s}}+\Delta^{{\rm v}{\rm w}}, \nonumber \\
&& Q^{\gamma_{\rm v}}-\Gamma^{\gamma_{\rm v}}\leq\sum_{n=0}^{\rm S_{cut}} p_n^{\gamma_{\rm v}} Y_n^{\gamma_{\rm s}}+\Delta^{{\rm w}{\rm s}}+\Delta^{{\rm v}{\rm w}}, \nonumber \\
&& Q^{\gamma_{\rm w}}\geq\sum_{n=0}^{\rm S_{cut}} p_n^{\gamma_{\rm w}} Y_n^{\gamma_{\rm s}}+\Delta^{{\rm w}{\rm s}}, \nonumber \\
&& Q^{\gamma_{\rm w}}-\Gamma^{\gamma_{\rm w}}\leq\sum_{n=0}^{\rm S_{cut}} p_n^{\gamma_{\rm w}} Y_n^{\gamma_{\rm s}}+\Delta^{{\rm w}{\rm s}}, \nonumber \\
&& 0\leq Y_n^{\gamma_{\rm s}}\leq 1, \ \forall n\leq {\rm S_{cut}}, \nonumber \\
&& -D_{{\rm w},{\rm s}}\leq \Delta^{{\rm w}{\rm s}}\leq D_{{\rm w},{\rm s}}, \quad -D_{{\rm v},{\rm w}}\leq \Delta^{{\rm v}{\rm w}}\leq D_{{\rm v},{\rm w}}, \nonumber \\
&& -D_{n,{\rm s,v,w}}\le\Delta^{{\rm w}{\rm s}}+q_{n{\rm vw}}\Delta^{{\rm v}{\rm w}}\le D_{n,{\rm s,v,w}}, \ \forall n\leq {\rm S_{cut}}, \nonumber \\
&& -D_{n,{\rm v,s,w}}\le q_{n{\rm sw}}\Delta^{{\rm w}{\rm s}}+\Delta^{{\rm v}{\rm w}}\le D_{n,{\rm v,s,w}}, \ \forall n\leq {\rm S_{cut}}, \nonumber \\
&& -D_{n,{\rm w,s,v}}\le
q_{n{\rm sv}}\Delta^{{\rm w}{\rm s}}-(1-q_{n{\rm sv}})\Delta^{{\rm v}{\rm w}}\le D_{n,{\rm w,s,v}}, \ \forall n\leq {\rm S_{cut}}
\end{eqnarray}
Note that the value of the parameters $D_{j,k}$ and $D_{n,j,k,l}$, with $j, k, l\in\{{\rm s, v, w}\}$, is provided in~\ref{appen_parD}. Also, the value of the observables
$Q^{\gamma_j}$ for a typical channel model can be found
in~\ref{toolbox}.
The linear program above has ${\rm S_{cut}}+3$ unknown parameters: $Y_n^{\gamma_{\rm s}}$, $\Delta^{{\rm w}{\rm s}}$ and $\Delta^{{\rm v}{\rm w}}$. Its solution is directly
$Y_{0{\rm L}}^{\gamma_{\rm s}}$.

\subsection{Estimation of $Y_{1{\rm L}}^{\gamma_{\rm s}}$}

To calculate $Y_{1{\rm L}}^{\gamma_{\rm s}}$, we can reuse the linear program given by Eq.~(\ref{eq_solqqb}), only substituting its linear objective function with $Y_1^{\gamma_{\rm s}}$.

\subsection{Estimation of $e_{1{\rm U}}^{\gamma_{\rm s}}$}

To obtain $e_{1{\rm U}}^{\gamma_{\rm s}}$, we can again reuse the linear program given by Eq.~(\ref{eq_solqqb}), only making the following three changes. First, all the parameters now refer to the X basis rather than the Z basis. For example, $Q^{\gamma_j}$ now denotes the gain when Alice selects the intensity setting $\gamma_j$ and both Alice and Bob use the X basis, and similarly for the other quantities that appear in Eq.~(\ref{eq_solqqb}).
Second, we substitute the parameters $Q^{\gamma_j}$ with $Q^{\gamma_j}E^{\gamma_j}$ for all $j\in\{{\rm s, v, w}\}$, and we replace the yields $Y_n^{\gamma_{\rm s}}$ with other variables that we will denote as $\omega_n^{\gamma_{\rm s}}$. These variables represent the value of $Y_n^{\gamma_{\rm s}}e_n^{\gamma_{\rm s}}$. Third, we substitute the linear objective function with $-\omega_1^{\gamma_{\rm s}}$, where the minus sign is because Eq.~(\ref{eq_solqqb}) is a minimisation problem and we are interested in obtaining an upper bound for $\omega_1^{\gamma_{\rm s}}$.

If we denote the solution to this optimisation problem as $n_{\rm sol}$, then $e_{1{\rm U}}^{\gamma_{\rm s}}$ is simply given by
\begin{equation}
e_{1{\rm U}}^{\gamma_{\rm s}}=-\frac{n_{\rm sol}}{Y_{1{\rm L}}^{\gamma_{\rm s}}},
\end{equation}
where, again, $Y_{1{\rm L}}^{\gamma_{\rm s}}$ now denotes a lower bound on the yield of the single-photon pulses when both Alice and Bob employ the X basis.
The value of the observables
$Q^{\gamma_j}$ and $E^{\gamma_j}$ for a typical channel model is provided
in~\ref{toolbox}.

\section{Estimation of $D_{n, j, k}$ and $D_{n, j, k,l}$}
\label{appen_parD}

In this Appendix we calculate the parameters $D_{n, j, k}$ and $D_{n, j, k,l}$
for the three examples studied in Sec.~\ref{simulations}. These parameters are needed to estimate a lower bound on
the yields $Y_0^{\gamma_{\rm s}}$ and $Y_1^{\gamma_{\rm s}}$, together with an upper bound on the phase error rate $e_1^{\gamma_{\rm s}}$, which is done in~\ref{app_est}.

All these examples correspond to individual THA, which implies that the states $\rho_{n, \gamma_{j}^{i}}$, which are accessible to Eve, do not depend on the instance $i$. In addition, they assume, as expected in most practical situations, that there is no correlation between Alice's system ${\rm A_p}$ and Eve's system ${\rm E_{p}'}$. That is, $\rho_{n, \gamma_{j}^{i}}={\hat P}(\ket{n})\otimes\rho_{\gamma_{j}}$. This means, in particular, that
\begin{eqnarray}\label{par_Ds}
D_{n, j, k}&=&d\left(\rho_{\gamma_{j}}, \rho_{\gamma_{k}}\right), \nonumber \\
D_{n, j, k,l}&=&d\left(\rho_{\gamma_{j}},
q_{nkl}\rho_{\gamma_{k}}+(1-q_{nkl})\rho_{\gamma_{l}}\right),
\end{eqnarray}
for all $n$. In this scenario, the parameters $D_{n, j, k}$ do not depend on the photon number $n$ and we will denote them as $D_{j, k}$. Next, we calculate these quantities for the different cases.

\subsection{Individual THA - Case~1:}\label{subsec:appen_parD_Case1}
In this example, the states $\rho_{\gamma_{j}}$ are of the form $\rho_{\gamma_{j}}={\hat P}(\ket{\beta e^{i\theta_{\gamma_j}}})$. We have, therefore, that
\begin{equation}
D_{j, k}=\sqrt{1-|\bra{\beta e^{i\theta_{\gamma_j}}}\beta e^{i\theta_{\gamma_k}}\rangle|^2}=\sqrt{1-e^{2\beta^2\left[\cos{(\theta_{\gamma_k}-\theta_{\gamma_j})}-1\right]}}.
\end{equation}
Here we can assume, without loss of generality, that $\theta_{\gamma_{\rm s}}=0$. Moreover, we will denote $\beta^2:=I_{\rm max}$. This implies, in particular, that
$D_{{\rm w, s}}$ and $D_{{\rm v, w}}$ are given by
\begin{eqnarray}\label{sun_morn}
D_{{\rm w, s}}&=&\sqrt{1-e^{2I_{\rm max}\left[\cos{(\theta_{\gamma_{\rm w}})}-1\right]}}, \nonumber \\
D_{{\rm v, w}}&=&\sqrt{1-e^{2I_{\rm max}\left[\cos{(\theta_{\gamma_{\rm w}}-\theta_{\gamma_{\rm v}})}-1\right]}}.
\end{eqnarray}

The parameters $D_{n, j, k,l}$ have the form
\begin{equation}
D_{n, j, k,l}=\frac{1}{2}\left|{\hat P}(\ket{\beta e^{i\theta_{\gamma_j}}})-q_{nkl}{\hat P}(\ket{\beta e^{i\theta_{\gamma_k}}})-(1-q_{nkl}){\hat P}(\ket{\beta e^{i\theta_{\gamma_l}}})\right|.
\end{equation}
In order to calculate these quantities we use the following Claim, which requires to obtain the eigenvalues of a $3\times3$ matrix.

\noindent {\bf Claim.} {\it Let $\{\ket{\alpha_i}\}_{i=\{1,2,3\}}$, be three normalised but not necessarily orthogonal vectors, and
let $\lambda_i$ be the eigenvalues of a $3\times3$ matrix $A$ defined as
\begin{equation}\label{juev}
(A)_{i,j}=\delta_{i,1}\bra{\alpha_1}\alpha_j\rangle-p\delta_{i,2}\bra{\alpha_2}\alpha_j\rangle-(1-p)\delta_{i,3}\bra{\alpha_3}\alpha_j\rangle,
\end{equation}
with $1\geq p\geq 0$ and where $\delta_{i,j}$ is the Kronecker delta. Then
\begin{equation}
\frac{1}{2}\left|{\hat P}(\ket{\alpha_1})-p{\hat P}(\ket{\alpha_2})-(1-p){\hat P}(\ket{\alpha_3})\right|=\frac{1}{2}\sum_i |\lambda_i|
\end{equation}
}

\begin{proof} To calculate the trace distance of $\rho:={\hat P}(\ket{\alpha_1})-p{\hat P}(\ket{\alpha_2})-(1-p){\hat P}(\ket{\alpha_3})$ we need
to determine its eigenvalues. Moreover, from the properties of the determinant we have that
${\rm Det}(\rho-\lambda{\hat {\mathbbm{1}}})={\rm Det}(V^{-1}\rho V-\lambda{\hat {\mathbbm{1}}})$ for any invertible linear operation $V$.
Then, we can construct $V$ as follows
\begin{eqnarray}
V\ket{i}&=&\ket{\alpha_i}, \quad \quad {\rm and}\quad \quad \bra{i}V^{-1}=\bra{\bar \alpha_i},
\end{eqnarray}
where $\{\ket{i}\}_{i=\{1,2,3\}}$ is an orthonormal basis, and $\ket{\bar \alpha_i}$ represent unnormalised vectors satisfying
$\bra{\bar \alpha_i}\alpha_j\rangle=\delta_{i,j}$. With these definitions, we can use $V$ and $\ket{\bar \alpha_i}$ to relate the matrix elements of
$V^{-1}\rho V$ defined in the orthogonal basis $\{\ket{i}\}_i$ to those of $\rho$ defined in the nonorthogonal basis $\{\ket{\alpha_i}\}_i$. In particular, we
have that
\begin{eqnarray}
\bra{i}V^{-1}\rho V\ket{j}&=&\bra{\bar \alpha_i}\rho\ket{\alpha_j}=\bra{\bar \alpha_i}{\hat P}(\ket{\alpha_1})\ket{\alpha_j}-p\bra{\bar \alpha_i}{\hat P}(\ket{\alpha_2})\ket{\alpha_j}
\nonumber \\
&-&(1-p)\bra{\bar \alpha_i}{\hat P}(\ket{\alpha_3})\ket{\alpha_j}=(A)_{i,j},
\end{eqnarray}
with $(A)_{i,j}$ given by Eq.~(\ref{juev}).
\end{proof}

\subsection{Individual THA - Case~2:}\label{subsec:appen_parD_Case2}
In this example, the states $\rho_{\gamma_{j}}$ are of the form $\rho_{\gamma_{j}}={\hat P}(\ket{\beta_{\gamma_j} e^{i\theta_{\gamma_j}}})$, where the
amplitudes $\beta_{\gamma_j}$ are given in Eq.~(\ref{set2}).
That is, here we assume that the back-reflected light that goes to Eve is attenuated in a similar manner as Alice's signals. In this scenario, we have that
\begin{eqnarray}
D_{j, k}&=&\sqrt{1-|\bra{\beta_{\gamma_j} e^{i\theta_{\gamma_j}}}\beta_{\gamma_k} e^{i\theta_{\gamma_k}}\rangle|^2} \nonumber \\
&=&\sqrt{1-e^{-\beta_{\gamma_j}^2-\beta_{\gamma_k}^2+2\beta_{\gamma_j}\beta_{\gamma_k}  \cos{(\theta_{\gamma_k}-\theta_{\gamma_j})} }}.
\end{eqnarray}
Again, if we assume, without loss of generality, that $\theta_{\gamma_{\rm s}}=0$ and we use Eq.~(\ref{set2}) we find that the
quantities $D_{{\rm w, s}}$ and $D_{{\rm v, w}}$ are given by
\begin{eqnarray}\label{sun_morn2}
D_{{\rm w, s}}&=&\sqrt{1-e^{-\frac{I_{\rm max}}{\gamma_{\rm s}}\left[\gamma_{\rm s}+\gamma_{\rm w}-2\sqrt{\gamma_{\rm s}\gamma_{\rm w}}\cos{(\theta_{\gamma_{\rm w}})}\right]}}, \nonumber \\
D_{{\rm v, w}}&=&\sqrt{1-e^{-\frac{I_{\rm max}}{\gamma_{\rm s}}\left[\gamma_{\rm v}+\gamma_{\rm w}-2\sqrt{\gamma_{\rm v}\gamma_{\rm w}}\cos{(\theta_{\gamma_{\rm w}}-\theta_{\gamma_{\rm v}})}\right]}}.
\end{eqnarray}

The parameters $D_{n, j, k,l}$ have the form
\begin{eqnarray}
D_{n, j, k,l}&=&\frac{1}{2}\left|{\hat P}(\ket{\beta_{\gamma_j} e^{i\theta_{\gamma_j}}})-q_{nkl}{\hat P}(\ket{\beta_{\gamma_k} e^{i\theta_{\gamma_k}}})-(1-q_{nkl}){\hat P}(\ket{\beta_{\gamma_l} e^{i\theta_{\gamma_l}}})\right|. \nonumber \\
\end{eqnarray}
Like in the previous subsection, we calculate these quantities by using the Claim introduced above.

\subsection{Individual THA - Case~3:}\label{subsec:appen_parD_Case3}
Here the states $\rho_{\gamma_{j}}$ are of the form given by Eq.~(\ref{set3})
with the intensities $\beta_{\gamma_j}^2$ given by Eq.~(\ref{set2}). This corresponds to the scenario where Eve's back-reflected light is phase-randomised and, moreover, it is attenuated in a similar manner as Alice's signals. In this situation, we have that
\begin{eqnarray}\label{fri}
D_{j, k}&=&\frac{1}{2}\sum_{n=0}^\infty\left\vert e^{-\beta_{\gamma_j}^2}\frac{\beta_{\gamma_j}^{2n}}{n!}-e^{-\beta_{\gamma_k}^2}\frac{\beta_{\gamma_k}^{2n}}{n!}\right\vert.
\end{eqnarray}
This means, in particular, that the parameters $D_{{\rm w, s}}$ and $D_{{\rm v, w}}$ are given by
\begin{eqnarray}\label{sun_morn3}
D_{{\rm w, s}}&=&\frac{e^{-I_{\rm max}}}{2}\sum_{n=0}^\infty \frac{I_{\rm max}^n}{n!}\left\vert1-e^{I_{\rm max}(1-\gamma_{\rm w}/\gamma_{\rm s})}\left(\frac{\gamma_{\rm w}}{\gamma_{\rm s}}\right)^n\right\vert, \nonumber \\
D_{{\rm v, w}}&=&\frac{1}{2}e^{-\frac{I_{\rm max}\gamma_{\rm v}}{\gamma_{\rm s}}}\sum_{n=0}^\infty \frac{(I_{\rm max}\gamma_{\rm v}/\gamma_{\rm s})^n}{n!}\left\vert1-e^{I_{\rm max}\gamma_{\rm v}/\gamma_{\rm s}(1-\gamma_{\rm w}/\gamma_{\rm v})}\left(\frac{\gamma_{\rm w}}{\gamma_{\rm v}}\right)^n\right\vert.
\end{eqnarray}
These expressions involve an infinite number of terms. However, one can easily upper bound them with a finite sum. For instance, it can be shown that
when $I_{\rm max}\leq \log{2}$ and $\gamma_{\rm s}\geq \gamma_{\rm v}\geq \gamma_{\rm w}$ (which is always satisfied in the simulation results shown in Sec.~\ref{simulations}) $D_{{\rm w, s}}$ and $D_{{\rm v, w}}$ can be upper bounded as
\begin{eqnarray}\label{brecol}
D_{{\rm w, s}}&\leq&\frac{1}{2}-\frac{e^{-I_{\rm max}}}{2}\sum_{n=0}^{P_{\rm cut}} \frac{I_{\rm max}^n}{n!}\left[1-\left\vert1-e^{I_{\rm max}(1-\gamma_{\rm w}/\gamma_{\rm s})}\left(\frac{\gamma_{\rm w}}{\gamma_{\rm s}}\right)^n\right\vert\right], \nonumber \\
D_{{\rm v, w}}&\leq&\frac{1}{2}-\frac{1}{2}e^{-\frac{I_{\rm max}\gamma_{\rm v}}{\gamma_{\rm s}}}\sum_{n=0}^{P_{\rm cut}} \frac{(I_{\rm max}\gamma_{\rm v}/\gamma_{\rm s})^n}{n!}
\nonumber \\
&\times&
\left[1-\left\vert1-e^{I_{\rm max}\gamma_{\rm v}/\gamma_{\rm s}(1-\gamma_{\rm w}/\gamma_{\rm v})}
\left(\frac{\gamma_{\rm w}}{\gamma_{\rm v}}\right)^n\right\vert\right].
\end{eqnarray}
for any $P_{\rm cut}\geq 1$. To see this, let us consider, for instance, the parameter $D_{{\rm w, s}}$. From Eq.~(\ref{sun_morn3}) we have that
$D_{{\rm w, s}}$ satisfies
\begin{eqnarray}\label{sun_morn4}
D_{{\rm w, s}}&=&\frac{e^{-I_{\rm max}}}{2}\sum_{n=0}^{P_{\rm cut}} \frac{I_{\rm max}^n}{n!}\left\vert1-e^{I_{\rm max}(1-\gamma_{\rm w}/\gamma_{\rm s})}\left(\frac{\gamma_{\rm w}}{\gamma_{\rm s}}\right)^n\right\vert+\frac{e^{-I_{\rm max}}}{2}\sum_{n=P_{\rm cut}+1}^\infty \frac{I_{\rm max}^n}{n!}, \nonumber \\
&\times&\left\vert1-e^{I_{\rm max}(1-\gamma_{\rm w}/\gamma_{\rm s})}\left(\frac{\gamma_{\rm w}}{\gamma_{\rm s}}\right)^n\right\vert
\le \frac{e^{-I_{\rm max}}}{2}\sum_{n=0}^{P_{\rm cut}} \frac{I_{\rm max}^n}{n!}\nonumber \\
&\times&\left\vert1-e^{I_{\rm max}(1-\gamma_{\rm w}/\gamma_{\rm s})}\left(\frac{\gamma_{\rm w}}{\gamma_{\rm s}}\right)^n\right\vert+\frac{e^{-I_{\rm max}}}{2}\sum_{n=P_{\rm cut}+1}^\infty \frac{I_{\rm max}^n}{n!}.
\end{eqnarray}
In the inequality condition we have used the fact that
\begin{equation}
\left\vert1-e^{I_{\rm max}(1-\gamma_{\rm w}/\gamma_{\rm s})}\left(\frac{\gamma_{\rm w}}{\gamma_{\rm s}}\right)^n\right\vert\le 1
\end{equation}
for all $n\geq 0$ and $I_{\rm max}\leq \log{2}$ given that $\gamma_{\rm s}\geq \gamma_{\rm v}\geq \gamma_{\rm w}$. This is so because $e^{I_{\rm max}}\geq e^{I_{\rm max}(1-\gamma_{\rm w}/\gamma_{\rm s})}\left(\frac{\gamma_{\rm w}}{\gamma_{\rm s}}\right)^n\geq 0$. Finally, by substituting the term $e^{-I_{\rm max}}/2\sum_{n=P_{\rm cut}+1}^\infty I_{\rm max}^n/n!$
with $1/2[1-e^{-I_{\rm max}}\sum_{n=0}^{P_{\rm cut}} I_{\rm max}^n/n!]$ we obtain Eq.~(\ref{sun_morn3}). The derivation of the upper bound for $D_{{\rm v, w}}$ is analogous.

The parameters $D_{n, j, k, l}$ are given by
\begin{eqnarray}
D_{n, j, k, l}&=&\frac{1}{2}
\sum_{n=0}^\infty\left\vert e^{-\beta_{\gamma_j}^2}\frac{\beta_{\gamma_j}^{2n}}{n!}-q_{nkl}e^{-\beta_{\gamma_k}^2}\frac{\beta_{\gamma_k}^{2n}}{n!}
-(1-q_{nkl})e^{-\beta_{\gamma_l}^2}\frac{\beta_{\gamma_l}^{2n}}{n!}\right\vert.
\end{eqnarray}
If we substitute the intensities $\beta_{\gamma_j}^2$ with the values given in Eq.~(\ref{set2}) we have, therefore, that
\begin{eqnarray}\label{wed_morn}
D_{n,{\rm s, v, w}}&=&\frac{1}{2}
\sum_{n=0}^\infty e^{-I_{\rm max}}\frac{I_{\rm max}^n}{n!}\Bigg\vert
1-q_{n{\rm vw}}e^{I_{\rm max}(1-\gamma_{\rm v}/\gamma_{\rm s})}\left(\frac{\gamma_{\rm v}}{\gamma_{\rm s}}\right)^n \nonumber \\
&-&(1-q_{n{\rm vw}})e^{I_{\rm max}(1-\gamma_{\rm w}/\gamma_{\rm s})}\left(\frac{\gamma_{\rm w}}{\gamma_{\rm s}}\right)^n\Bigg|, \nonumber \\
D_{n,{\rm v, s, w}}&=&\frac{1}{2}
\sum_{n=0}^\infty e^{-I_{\rm max}}\frac{I_{\rm max}^n}{n!}\Bigg\vert
e^{I_{\rm max}(1-\gamma_{\rm v}/\gamma_{\rm s})}\left(\frac{\gamma_{\rm v}}{\gamma_{\rm s}}\right)^n-q_{n{\rm sw}} \nonumber \\
&-&(1-q_{n{\rm sw}})
e^{I_{\rm max}(1-\gamma_{\rm w}/\gamma_{\rm s})}\left(\frac{\gamma_{\rm w}}{\gamma_{\rm s}}\right)^n
\Bigg|, \nonumber \\
D_{n,{\rm w, s, v}}&=&\frac{1}{2}
\sum_{n=0}^\infty e^{-I_{\rm max}}\frac{I_{\rm max}^n}{n!}\Bigg\vert
e^{I_{\rm max}(1-\gamma_{\rm w}/\gamma_{\rm s})}\left(\frac{\gamma_{\rm w}}{\gamma_{\rm s}}\right)^n-q_{n{\rm sv}} \nonumber \\
&-&(1-q_{n{\rm sv}})
e^{I_{\rm max}(1-\gamma_{\rm v}/\gamma_{\rm s})}\left(\frac{\gamma_{\rm v}}{\gamma_{\rm s}}\right)^n
\Bigg|.
\end{eqnarray}
Again, these equations involve an infinite number of terms. However, as above,
it can be shown that when $I_{\rm max}\leq \log{2}$ and $\gamma_{\rm s}\geq \gamma_{\rm v}\geq \gamma_{\rm w}$ the parameters $D_{n,{\rm s, v, w}}$, $D_{n,{\rm v, s, w}}$ and $D_{n,{\rm w, s, v}}$ are upper
bounded by
\begin{eqnarray}\label{qwe}
D_{n,{\rm s, v, w}}&\leq&\frac{1}{2}\Bigg\{1-\sum_{n=0}^{P_{\rm cut}} e^{-I_{\rm max}}\frac{I_{\rm max}^n}{n!}\Bigg(1-\Bigg\vert
1-q_{n{\rm vw}}e^{I_{\rm max}(1-\gamma_{\rm v}/\gamma_{\rm s})}\nonumber \\
&\times&\left(\frac{\gamma_{\rm v}}{\gamma_{\rm s}}\right)^n-
(1-q_{n{\rm vw}})e^{I_{\rm max}(1-\gamma_{\rm w}/\gamma_{\rm s})}\left(\frac{\gamma_{\rm w}}{\gamma_{\rm s}}\right)^n\Bigg|\Bigg)\Bigg\}, \nonumber \\
D_{n,{\rm v, s, w}}&\leq&\frac{1}{2}\Bigg\{q_{n{\rm sw}}+(1-q_{n{\rm sw}})e^{I_{\rm max}}
-\sum_{n=0}^{P_{\rm cut}} e^{-I_{\rm max}}\frac{I_{\rm max}^n}{n!}\Bigg(q_{n{\rm sw}}+(1-q_{n{\rm sw}})\nonumber \\
&\times&e^{I_{\rm max}}-\Bigg\vert
e^{I_{\rm max}(1-\gamma_{\rm v}/\gamma_{\rm s})}
\left(\frac{\gamma_{\rm v}}{\gamma_{\rm s}}\right)^n-q_{n{\rm sw}}
-(1-q_{n{\rm sw}})
e^{I_{\rm max}(1-\gamma_{\rm w}/\gamma_{\rm s})}\nonumber \\
&\times&
\left(\frac{\gamma_{\rm w}}{\gamma_{\rm s}}\right)^n
\Bigg|\Bigg)\Bigg\}, \nonumber \\
D_{n,{\rm w, s, v}}&\leq&\frac{1}{2}\Bigg\{q_{n{\rm sv}}+(1-q_{n{\rm sv}})e^{I_{\rm max}}
-\sum_{n=0}^{P_{\rm cut}} e^{-I_{\rm max}}\frac{I_{\rm max}^n}{n!}\Bigg(q_{n{\rm sv}}+(1-q_{n{\rm sv}})\nonumber \\
&\times&e^{I_{\rm max}}-\Bigg\vert
e^{I_{\rm max}(1-\gamma_{\rm w}/\gamma_{\rm s})}
\left(\frac{\gamma_{\rm w}}{\gamma_{\rm s}}\right)^n-q_{n{\rm sv}}
-(1-q_{n{\rm sv}})
e^{I_{\rm max}(1-\gamma_{\rm v}/\gamma_{\rm s})}\nonumber \\
&\times&
\left(\frac{\gamma_{\rm v}}{\gamma_{\rm s}}\right)^n
\Bigg|\Bigg)\Bigg\},
\end{eqnarray}
for any $P_{\rm cut}\geq 1$. To see this, the procedure is analogous to the one used
to derive Eq.~(\ref{brecol}).
In particular, let us consider the quantity $D_{n,{\rm s, v, w}}$. From Eq.~(\ref{wed_morn}) we have that $D_{n,{\rm s, v, w}}$ can be upper bounded as
\begin{eqnarray}\label{wed_morn2}
D_{n,{\rm s, v, w}}
&\le& \frac{1}{2}
\sum_{n=0}^{P_{\rm cut}} e^{-I_{\rm max}}\frac{I_{\rm max}^n}{n!}\Bigg\vert
1-q_{n{\rm vw}}e^{I_{\rm max}(1-\gamma_{\rm v}/\gamma_{\rm s})}\left(\frac{\gamma_{\rm v}}{\gamma_{\rm s}}\right)^n \nonumber \\
&-&(1-q_{n{\rm vw}})e^{I_{\rm max}(1-\gamma_{\rm w}/\gamma_{\rm s})}\left(\frac{\gamma_{\rm w}}{\gamma_{\rm s}}\right)^n\Bigg|
+\frac{1}{2}\sum_{n=P_{\rm cut}+1}^{\infty} e^{-I_{\rm max}}\frac{I_{\rm max}^n}{n!}.
\end{eqnarray}
Here we have used the fact that
\begin{eqnarray}\label{juev2}
&&\Bigg\vert1-q_{n{\rm vw}}e^{I_{\rm max}(1-\gamma_{\rm v}/\gamma_{\rm s})}\left(\frac{\gamma_{\rm v}}{\gamma_{\rm s}}\right)^n
-(1-q_{n{\rm vw}})e^{I_{\rm max}(1-\gamma_{\rm w}/\gamma_{\rm s})}\left(\frac{\gamma_{\rm w}}{\gamma_{\rm s}}\right)^n\Bigg| \nonumber \\
&&\leq 1,
\end{eqnarray}
for all $n\geq 0$ and $I_{\rm max}\leq \log{2}$ given that $\gamma_{\rm s}\geq \gamma_{\rm v}\geq \gamma_{\rm w}$.
Eq.~(\ref{juev2})
 holds because $e^{I_{\rm max}}\geq e^{I_{\rm max}(1-\gamma_{\rm k}/\gamma_{\rm s})}(\gamma_{\rm k}/\gamma_{\rm s})^n\geq 0$ for all $k\in\{{\rm v,w}\}$. Finally, by
 replacing in Eq.~(\ref{wed_morn2})
 $e^{-I_{\rm max}}/2\sum_{n=P_{\rm cut}+1}^{\infty} I_{\rm max}^n/n!$ with $1/2[1-\sum_{n=0}^{P_{\rm cut}} e^{-I_{\rm max}}I_{\rm max}^n/n!]$ one obtains
Eq.~(\ref{qwe}). The upper bounds for $D_{n,{\rm v, s, w}}$ and $D_{n,{\rm w, s, v}}$ can be obtained in a similar manner.

\section{Toolbox for Alice and Bob, and channel model}\label{toolbox}

In this Appendix we introduce a simple mathematical model to characterise Alice's and Bob's devices, together with the behaviour of a typical quantum channel. This model is used to simulate the observed experimental data $Q^{\gamma_j}$ and $E^{\gamma_j}$, with $j\in\{{\rm s, v, w}\}$, which is needed to evaluate the examples considered in Sec.~\ref{simulations}. Here we will consider that $Q^{\gamma_j}$ and $E^{\gamma_j}$ do not depend on the basis setting, {\it i.e.},
they
are equal for both the Z and the X basis.

In particular, we assume the standard decoy-state BB84 protocol with phase-encoding. In each time slot, Alice prepares two WCP, the signal and the reference pulse, whose joint phase is perfectly randomised. Then, she selects at random a phase modulation $\phi\in\{0, \pi/2, \pi, 3\pi/2\}$ and applies it
to the signal pulse. The values $0$ and $\pi$ ($\pi/2$ and $3\pi/2$) correspond to the Z (X) basis. In addition, Alice uses an intensity modulator to randomly choose the intensity $\gamma\in\{\gamma_{\rm s}, \gamma_{\rm v}, \gamma_{\rm w}\}$ of both the signal and the reference pulse following the prescriptions of the decoy-state method. As a result, Alice sends Bob states of the form
\begin{equation}
\ket{\Psi^{\phi,\gamma}}_{\rm A_p}=\frac{1}{2\pi}\int_{0}^{2\pi} {\hat P}\left(\ket{\sqrt{\gamma^i}e^{i\theta}}_{\rm r}\ket{\sqrt{\gamma^i}e^{i(\theta+\phi)}}_{\rm s}\right){\rm d}\theta,
\end{equation}
where the subscript s (r) identifies the signal (reference) pulse and $\theta\in[0,2\pi)$ is a random phase.

On the receiving side, Bob uses a Mach-Zehnder interferometer to divide the incoming pulses into two possible paths. Then he applies a phase shift $\phi\in\{0,\pi/2\}$ together with a one-pulse delay to one of them, and he recombines both pulses at a $50:50$ beamsplitter. This beamsplitter has on its ends two single-photon detectors, which we denote as
$D_0$ and $D_1$. Whenever the relative phase between the two interfering pulses is $0$ ($\pm\pi$) only the detector  ${\rm D}_0$ (${\rm D}_1$) can produce a ``click'', which indicates that at least one photon has been detected.
In case that both detectors ``click'' Bob uses the standard post-processing step where he assigns a random value to the raw bit~\cite{norbert}.
Given that both detectors have the same quantum efficiency and assuming for the moment that there is no side-channel in Bob's measurement unit,
this data post-processing guarantees the so-called basis independent detection efficiency condition. That is, Bob's detection efficiency is
the same for both BB84 bases.
Each detector is described by a positive operator value measure
with two elements, ${\hat F}_{\rm no click}$ and ${\hat F}_{\rm click}$.
The outcome of ${\hat F}_{\rm no click}$ corresponds to a ``no click'' event, whereas the operator ${\hat F}_{\rm click}$ gives one detection ``click''.
These operators are given by
\begin{eqnarray}
{\hat F}_{\rm no click}&=&(1-p_{\rm d})\sum_{n=0}^\infty (1-\eta_{\rm det})^n{\hat P}(\ket{n}), \nonumber \\
{\hat F}_{\rm click}&=&{\hat {\mathbbm{1}}}-{\hat F}_{\rm no click}.
\end{eqnarray}
Here $p_{\rm d}$ denotes the detector's dark count rate and $\eta_{\rm det}$ is its detection efficiency.

The quantum channel introduces loss that can be parametrised by the transmission efficiency $\eta_{\rm channel}$ given by
\begin{equation}
\eta_{\rm channel}=10^{\frac{-\alpha d}{10}},
\end{equation}
where $\alpha$ is the loss coefficient of the channel measured in dB/km and $d$ is the transmission distance measured in km.
In addition, we assume that
the QKD setup has an intrinsic error rate  $e_{\rm d}$
due to misalignment and instability of the optical system.

By using the mathematical models above, it can be shown that the gain $Q^{\gamma_j}$ and the error rate $E^{\gamma_j}$ can be expressed as
\begin{eqnarray}
Q^{\gamma_j}&=&1-(1-p_{\rm d})^2 e^{-\gamma_j\eta_{\rm sys}}. \nonumber \\
E^{\gamma_j}&=&\frac{1}{2}+\frac{1}{2 Q^{\gamma_j}}(1-p_{\rm d})\left[e^{-\gamma_j\eta_{\rm sys}(1-e_{\rm d})}-e^{-\gamma_j\eta_{\rm sys}e_{\rm d}}\right],
\end{eqnarray}
where $\eta_{\rm sys}$ represents the overall loss of the system. It is given by
\begin{equation}
\eta_{\rm sys}=\eta_{\rm channel}\eta_{\rm B}\eta_{\rm det},
\end{equation}
with $\eta_{\rm B}$ being the internal loss of Bob's measurement device without considering
his detectors. That is, we assume that the total loss within Bob's receiver is $\eta_{\rm B}\eta_{\rm det}$.

\section{Approaching the optimal THA with phase-randomised coherent states}
\label{lastq}

In this Appendix we consider the scenario where Eve's back-reflected light is phase-randomised ({\it i.e.}, case 3 in Sec.~\ref{simulations}), and we analyse an alternative strategy for Eve. More precisely, we assume that Eve sends Alice $n$-photon Fock states instead of coherent states. This constitutes her optimal strategy in this situation, and below we
analyse how much could now the parameters $D_{j,k}$ deviate from the ones obtained in~\ref{appen_parD}.

Let us consider here the standard model of a beamsplitter with transmissivity $\eta_{\gamma_j}$ to characterise the loss introduced by Alice's device on Eve's input signals. Then, if Eve injects an $n$-photon state $\ket{n}$ into Alice's device, the state of the back-reflected light is given by
\begin{eqnarray}
\sigma_{\gamma_{j}}=\sum_{k=0}^\infty {n \choose k} \eta_{\gamma_j}^{k}(1-\eta_{\gamma_j})^{n-k}\ket{k}\bra{k},
\end{eqnarray}
The trace distance between these states and the ones given by Eq.~(\ref{set3}) is
\begin{eqnarray}\label{poi}
d(\rho_{\gamma_{j}}, \sigma_{\gamma_{j}})&:=&\frac{1}{2}\sum_{k=0}^{\infty}\left|P_{\eta_{\gamma_j}\mu}(k)-B_{\eta_{\gamma_j}}(n,k)\right|\,,
\end{eqnarray}
where
$P_{\eta_{\gamma_j}\mu}(k):=e^{-\eta_{\gamma_j}\mu}(\eta_{\gamma_j}\mu)^k/k!$ is a Poisson distribution of mean $\eta_{\gamma_j}\mu$,
$B_{\eta_{\gamma_j}}(n,k):={n \choose k} \eta_{\gamma_j}^{k}(1-\eta_{\gamma_j})^{n-k}$ is a Binomial distribution, and
$\eta_{\gamma_j}$ denotes Alice's transmission rate. When compared to the notation used in Eq.~(\ref{set3}), note that
$\beta_{\gamma_j}^2:=\eta_{\gamma_j}\mu$, with $\mu$ being the intensity of Eve's input pulses. Importantly,
from~\cite{Le1965} we have that whenever $\mu=n$ ({\it i.e.}, the intensity of Eve's input pulses is the same in both scenarios)
then Eq.~(\ref{poi}) can be upper bounded by
\begin{eqnarray}\label{poi2}
d(\rho_{\gamma_{j}}, \sigma_{\gamma_{j}})&\le&2\eta_{\gamma_j}.
\end{eqnarray}

Then, by using the triangle inequality we have that the trace distance between $\sigma_{\gamma_{j}}$ and $\sigma_{\gamma_{k}}$ is upper bounded by
\begin{eqnarray}
d(\sigma_{\gamma_{j}},\sigma_{\gamma_{k}})&=&\frac{1}{2}\sum_{k=0}^{\infty}\left|B_{\eta_{\gamma_j}}(n,k)-B_{\eta_{\gamma_k}}(n,k)\right| \le
2(\eta_{\gamma_j}+\eta_{\gamma_k})+D_{j,k},
\end{eqnarray}
where $D_{j,k}=d(\rho_{\gamma_j},\rho_{\gamma_k})$ is given by Eq.~(\ref{fri}).

In the examples considered in Sec.~\ref{simulations} the parameters $\eta_{\gamma_j}$ are typically very small (of the order of $10^{-13}-10^{-18}$ for a 1~GHz-clocked QKD system) for all $j\in\{{\rm s, v, w}\}$.
This means, in particular, that $d(\sigma_{\gamma_{j}},\sigma_{\gamma_{k}})\approx D_{j,k}$ and, therefore, the results presented in Sec.~\ref{simulations} (see case 3) are also valid for the scenario where Eve injects $n$-photon Fock states into Alice's device.

\section*{References}

\bibliographystyle{iopart-num}

\end{document}